\documentclass[aps,prb,twocolumn,superscriptaddress,longbibliography,noeprint]{revtex4-2}

\usepackage{amsmath}
\usepackage{graphicx}
\graphicspath{{./figures/}}
\usepackage{bm}
\usepackage{multirow}
\usepackage[table]{xcolor}
\usepackage[export]{adjustbox}
\makeatletter
\let\old@makecaption=\@makecaption
\usepackage{subcaption}
\let\@makecaption=\old@makecaption
\makeatother
\usepackage{color}
\usepackage{orcidlink}
\usepackage{xspace}
\usepackage{natbib}
\usepackage{siunitx}
\usepackage{miller}

\usepackage[version=3]{mhchem}

\newcommand{\bfa}{BaFe$_{2}$As$_2$\xspace}

\newcommand{\lfa}{LiFeAs\xspace}

\newcommand{\nfa}{NaFeAs\xspace}

\newcommand{\lfaofx}{LaFeAsO$_{1-x}$F$_{x}$\xspace}

\newcommand{\bs}[1]{\boldsymbol{#1}}
\renewcommand{\theta}{\vartheta}
\renewcommand{\phi}{\varphi}

\newcommand{\spm}{$s^{\pm}$\xspace}
\newcommand{\spp}{$s^{++}$\xspace}

\newcommand{\etal}{\textit{et al.}\xspace}

\newcommand{\tc}{$T_c$\xspace}

\begin{document}
	
	\title{Lattice Dynamics of LiFeAs studied by Inelastic Neutron Scattering and Density Functional Theory calculations}

\author{Akshay Tewari\,\orcidlink{0000-0002-9633-1490}}
\email{tewari@ph2.uni-koeln.de}
\affiliation{II.\, Physikalisches Institut, Universit\"at zu K\"oln, Z\"ulpicher Str.\ 77, 50937 K\"oln, Germany}
\author{Navid Qureshi\,\orcidlink{0000-0002-0727-2258}}
\affiliation{II.\, Physikalisches Institut, Universit\"at zu K\"oln, Z\"ulpicher Str.\ 77, 50937 K\"oln, Germany}
\affiliation{Institut Laue Langevin, BP156X, 38042 Grenoble Cedex, France}
\author{Rolf Heid}
\affiliation{Institut für Quantenmaterialien und Technologien, Karlsruher Institut für Technologie (KIT), Postfach 3640, D-76121 Karlsruhe, Germany}
\author{Andrea Piovano}
\affiliation{Institut Laue Langevin, BP156X, 38042 Grenoble Cedex, France}
\author{Yvan Sidis}
\affiliation{Universit\'e Paris-Saclay, CNRS, CEA, Laboratoire L\'eon Brillouin, 91191 Gif-sur-Yvette, France}
\affiliation{Institut Jean Lamour, CNRS UMR 7198, Universit\'e de Lorraine, Vandoeuvre-l\`es-Nancy F-54506, France}
\author{Luminita Harnagea}
\affiliation{Leibniz-Institut für Festk\"orper u. Werkstoffforschung (IFW) Dresden, Helmholtzstrasse 20, 01069 Dresden, Germany}
\affiliation{I-HUB Quantum Technology Foundation, Indian Institute of Science Education and Research, Pune 411008, India}
\author{Sabine Wurmehl}
\affiliation{Leibniz-Institut für Festk\"orper u. Werkstoffforschung (IFW) Dresden, Helmholtzstrasse 20, 01069 Dresden, Germany}
\author{Bernd B\"uchner}
\affiliation{Leibniz-Institut für Festk\"orper u. Werkstoffforschung (IFW) Dresden, Helmholtzstrasse 20, 01069 Dresden, Germany}
\author{Markus Braden\,\orcidlink{0000-0002-9284-6585}}
\email{braden@ph2.uni-koeln.de}
\affiliation{II.\, Physikalisches Institut, Universit\"at zu K\"oln, Z\"ulpicher Str.\ 77, 50937 K\"oln, Germany}

\begin{abstract}

We investigated the lattice dynamics of the unconventional superconductor LiFeAs using inelastic neutron scattering experiments and density-functional theory (DFT) calculations. By comparing the neutron scattering intensities with lattice-dynamics simulations we can identify the polarization symmetry of all modes along the main-symmetry directions yielding a complete experimental picture of the phonon dispersion. 
Overall there is good agreement between the experimental and DFT results, which renders an overlooked strong electron phonon coupling unlikely. Our DFT calculations reveal only a small averaged electron-phonon coupling constant. The transversal acoustic in-plane branches exhibit a normal dispersion for small propagation vectors indicating the absence of a nematic instability.
Several modes exhibit considerable hardening upon cooling that can be attributed to the anisotropic shrinking of the 
LiFeAs lattice.  

\end{abstract}

\date{\today}

\maketitle

\section{Introduction}

The discovery of superconductivity at 26\,K in \lfaofx \cite{Kamihara2008} opened the path to the new class of Fe-based high-temperature superconductors \cite{Hosono2015}. The materials of this broad family are distinguished by their chemical
formula: 1111 (eg. $RE$FeAsO \cite{Kamihara2008,lfao2}), 122 (eg. $A$Fe$_2$As$_2$ with $A$=Sr, Ba \cite{bfa1,bfa2}), 111 (eg. $A$FeAs \cite{lfa1,lfa2} with $A$=Li, Na), 11 (eg. Fe$X$ with X=S, Se, Te \cite{fc1,fc2}). Superconductivity can be induced or altered in these compounds via electron or hole doping \citep{doping1,doping2}, isovalent substitution \cite{substitution1} or pressure \citep{pressure1,Hosono2015}. Theoretical analyses of the electron-phonon coupling yielded only weak interaction and rendered a phonon-driven pairing unlikely \cite{Boeri2008,Haule2008} in agreement with the finding of a weak and inverse isotope effect \cite{Shirage2009}. 
Instead, unconventional superconductivity was proposed to emerge through magnetic interaction resulting in a \spm symmetry of the superconducting order parameter with phase changes between different sheets of the Fermi surface \cite{Mazin2008,Chubukov2008,Hirschfeld2011,Scalapino2012,Chubukov2008,Hirschfeld2016}. In this scenario, strong nesting between hole and electron pockets drives antiferromagnetic (AFM) order in the parent compounds and the associated AFM fluctuations mediate superconducting pairing. The emergence of strong spin-resonance modes in the superconducting state and the phase diagrams with neighboring or even competing AFM and superconducting phases give strong support for this spin-fluctuation mediated coupling mechanism \cite{Dai2015a,Scalapino2012}.
Spin-orbit coupling results in spin-space anisotropy of the magnetic excitations
in the parent compounds \cite{QureshiBraden2012,Wang2013} as well as in the superconducting materials, in which split and anisotropic spin-resonance modes emerge \cite{Steffens2013}.  
Including the spin-orbit interaction on a random-phase-approximation level, Scherer and Andersen qualitatively explain the experimental observations of anisotropy \cite{Scherer2018,Scherer2019,Scherer2019b} underlining the reliability of this approach.
The magnetic ordering,
which breaks the tetragonal symmetry of the ideal layers, is associated with an electronic ordering that frequently precedes the onset of magnetic order. Since there is strong evidence that the orthorhombic strain is not the primary order parameter of this transition, the state is called nematic (a breaking of rotational symmetry without loss of translation symmetry) \cite{Wang2021,Boehmer2022,Fernandes2022}. However, the possible role of nematicity in the superconducting pairing remains an open issue. The multiorbital character of the Fe-based compounds has strong impact not only on the superconducting state but also on the normal state as the local Hund's interaction becomes prominent \cite{Haule2009}.

There are only a few materials that exhibit superconductivity in their pristine form including FeSe, CaKFe$_4$As$_4$ and LiFeAs \cite{Fernandes2022} and these attracted special attention already due to the lack of intrinsic disorder, which is particularly favorable for surface-sensitive studies. 
LiFeAs was synthesized in 1968 \cite{Juza1968} but only little investigated till the discovery of its superconductivity in 2008 \cite{Wang2008a,Tapp2008}.
The critical temperature, \tc , values range between 16\,K and 18\,K in the literature \cite{Heyer2011,Li2013,Wang2008a,Hanaguri2012,Pitcher2010,Pitcher2008,Aswartham2011,Stockert2011a,Chi2012,Khim2011}
most likely due to the sensitivity on vacancies, and superconductivity is totally suppressed for $y$\,$\sim$\,2\% in \ce{Li$_{1-y}$Fe$_{1+y}$As} \cite{Pitcher2010}.
The sister compound \nfa exhibits the common tetragonal to orthorhombic phase transition at $\sim$\SI{60}{K} and AFM order at $\sim$\SI{45}{K} \cite{Parker2010,Wang2012}, in contrast \lfa remains tetragonal and paramagnetic to the lowest temperatures \cite{Tapp2008,Wang2008a,Pitcher2008}.
In other Fe-based superconductors \tc typically emerges or increases by applying pressure or doping, while  \tc only diminishes in \lfa . 
Upon electron doping induced by substitution with Co, superconductivity completely vanishes at $x \sim \SI{12}{\%}$ \cite{Aswartham2011,Ye2014,Dai2015}, while
hole doping by substituting with V rapidly decreases \tc at a rate of 7 K per 1\% V accompanied by formation of high magnetic moments \cite{Xing2016}. 
External pressure also reduces \tc by approximately -1.4\,K/GPa \cite{Zhang2009,Gooch2009}.

LiFeAs exhibits facile cleavage, yielding surfaces with high purity and good properties that were widely explored by angle-resolved photoemission spectroscopy (ARPES) \cite{Borisenko2010,Umezawa2012,Ye2014,Miao2014,Miao2015,Li2020,Kushnirenko2020,Day2022} and scanning tunneling microscopy \cite{Hess2013,Chi2012}.
The Fermi surface of LiFeAs consists of two hole-like pockets centered at the $\Gamma$ point (0 0 $l$) in the Brillouin zone
and of two electron-like pockets at the M points ($\pm\frac{1}{2}$  0  $l$) or (0 $\pm\frac{1}{2}$ $l$)
but there is sizable dispersion vertical to the layers \cite{Day2022}.
\lfa shows topological surface states and is therefore a candidate for topological superconductivity \cite{Zhang2019,Day2022}.
Early ARPES studies on \lfa  by  Borisenko \etal \cite{Borisenko2010} revealed only poor nesting conditions between
the hole and electron pockets of the Fermi surface, which were confirmed by many following studies \cite{Umezawa2012,Ye2014,Miao2014,Miao2015,Li2020,Kushnirenko2020,Day2022}. 
Therefore, the standard AFM-fluctuations-based \spm model for the superconducting pairing can be questioned for \lfa but 
a three-dimensional analysis proposed that the scattering processes between electron and hole pockets dominate
the pairing also in this material \cite{Wang2013e}.
Indeed, inelastic neutron scattering (INS) experiments find AFM fluctuations appearing at incommensurate propagation vectors $\bf{q}_{inc}$=(0.5$\pm\xi$,0.5$\mp\xi$,$l$) close to the commensurate  values, where AFM correlations emerge in the other FeAs materials \cite{Qureshi2012,Qureshi2014a,Knolle2012,Wang2012a}. However, these fluctuations are much weaker than those in Co-doped \bfa with similar \tc and there is only a gradual uptake at \SI{7.5}{meV} accompanied by an intensity decrease at low energies upon passing into the superconducting state \cite{Qureshi2012,Qureshi2014a} in contrast to the strong resonance excitations observed in 122 materials\,\cite{Dai2015a}.

\begin{figure}[t]
	\centering
	\includegraphics[width=0.46\columnwidth]{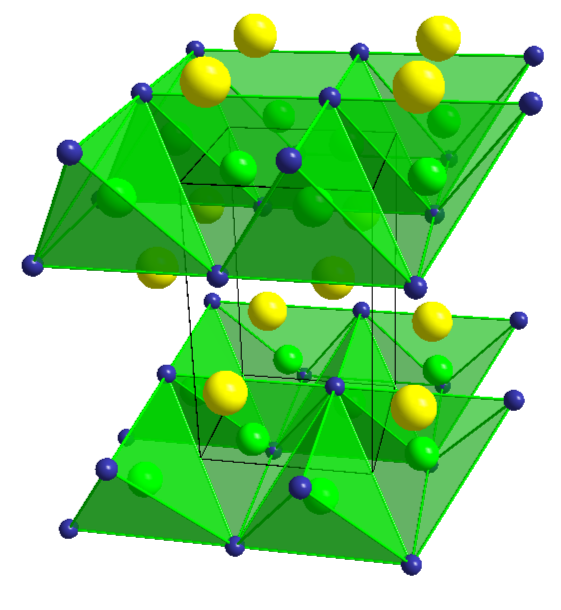}
	\includegraphics[width=0.525\columnwidth]{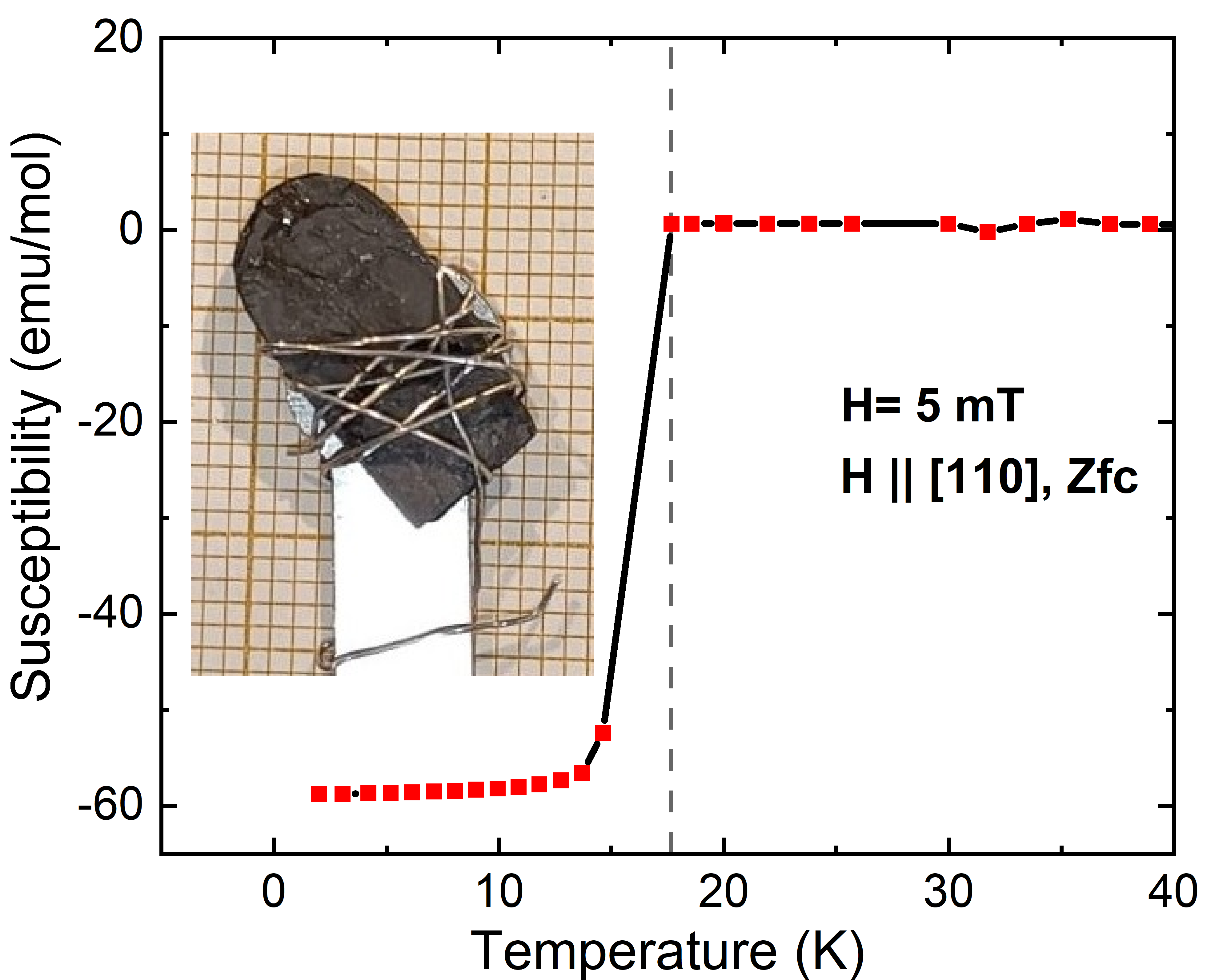}
	\caption{Crystal structure of LiFeAs with FeAs tetrahedron layers separated by planes of Li (in yellow) (a). The measurement of the magnetization at a magnetic field of 5\,mT parallel to [110]  on a part of a crystal with a SQUID magnetometer determines the superconducting transition temperature to $T_\mathrm{c}$\,=\,17.6(6)\,K (b). The magnetization data were taken after zero-field cooling. The inset shows the mounting of the large crystal for the INS experiments.}
	\label{LiFeAs_SQUID.png}
\end{figure}

The absence of strong nesting conditions inspired the proposal of pairing through ferromagnetic fluctuations leading 
to a $p$-wave superconducting order parameter \cite{Brydon2011}.
A non-vanishing Knight-shift in an NMR experiment \cite{Baek2012} and  quasi-particle interference (QPI) patterns \cite{Haenke2012} supported such a scenario, but a polarized neutron diffraction study of the spin susceptibility finds the unambiguous suppression of electron susceptibility expected for spin-singlet pairing \cite{Brand2014}.
The discussion about the superconducting pairing mechanism and symmetry is thus more controversial for \lfa  than for the other families \cite{Wang2013b,Ahn2014a,Nourafkan2016,Allan2015} and also theories based on phonon-assisted orbital fluctuations were applied for \lfa yielding the sign-preserving \spp superconducting order-parameter symmetry \cite{Saito2014a,Saito2015}. 
Two high-resolution ARPES experiments on \lfa revealed three energy scales of strong electron-bosonic-mode coupling that can tentatively be
attributed to phonons \cite{Li2020,Kordyuk2011} rendering a quantitative study of the lattice dynamics in \lfa desirable.

Due to the reports of insufficiently strong electron-phonon coupling in the FeAs-based superconductors  \cite{Boeri2008,Haule2008} their lattice dynamics attracted less attention and only selected aspects
were studied mostly related with either the structural transitions or the nematicity \cite{Reznik2009,Zbiri2009,Lee2010,Fukuda2011,Parshall2014,Parshall2015,Weber2018,Merritt2020,Merritt2016}. In contrast, the lattice dynamics of many cuprate families were studied in great detail revealing the full phonon dispersion separated according to the different irreducible representations for phonon propagation vectors along the main symmetry directions \cite{Chaplot1995}. Such a complete and detailed picture has not been obtained for any of the FeAs-based superconductors. For \lfa some zone-center modes were studied by Raman scattering \cite{Um2012} and the temperature dependence of transversal acoustic modes close to the Brillouin zone center was investigated by INS \cite{Merritt2016}. In addition density functional theory (DFT) calculations of the phonon dispersion were reported by Jishi \etal \cite{jishi} and Huang \etal \cite{huang} that, however, only poorly agree with each other. 

We performed comprehensive INS studies on the phonon dispersion of \lfa using thermal triple-axis spectrometers (TAS) and analyzed the data by lattice-dynamical calculations with a force-constant model. In addition, we applied DFT 
to calculate the phonon dispersion.
By successively refining the model and by comparing the predicted phonon intensities with the INS measurements we  identify the character in terms of the phonon polarization and obtain an almost complete description of the phonon modes along the main-symmetry directions. Overall the experimental and theoretical DFT results agree well with each other suggesting
that the main characteristics of the lattice dynamics are well captured by the calculations.

\section{Experimental}

The single crystals used in this study were synthesized at IFW Dresden using the self-flux technique \cite{Morozov2010} and were already studied by INS experiments aiming at the magnetic excitations \cite{Qureshi2012,Qureshi2014a}. 
Recent magnetization measurements on a SQUID magnetometer revealed the superconducting transition temperature of 17.6(6)\,K on a small part cut from a large crystal, see Fig.\,\ref{LiFeAs_SQUID.png}.
Two large LiFeAs samples of 400 and 600\,mg weight were oriented in the  [110]/[001], [100]/[010] and [100]/[001] scattering geometries to access the transverse and longitudinal modes along the three high-symmetry paths  $\Delta$ along [100], $\Sigma$ along [110] and $\Lambda$ along [001]. 

The alkali-earth metal Li is very reactive, and, when exposed to moisture or air, decomposes into Lithium oxide and Lithium nitride. Hence the samples were always kept and manipulated under an Argon atmosphere in the glove box. The samples were mounted in a tight Al can, which was inserted in an ILL orange-type liquid Helium cryostat or in a closed-cycle cooler for the INS experiments. 

A first set of INS experiment was performed on the thermal TAS 1T at the Laboratoire L\'eon Brillouin in Saclay (France). The instrument uses double-focusing monochromator and analyzer crystals of highly oriented pyrolytic graphite (HOPG) with the (002) reflection. A HOPG filter in the scattered beam suppressed higher-order contaminations and the crystal was cooled in a closed-cycle refrigerator. The sample was oriented in the [100]/[010] scattering plane and scans were performed with a fixed final neutron momentum of $k_f$\,=\,2.66\,\AA$^{-1}$.

\begin{figure}[h]
	\begin{center}
		\includegraphics[width=0.9\columnwidth]{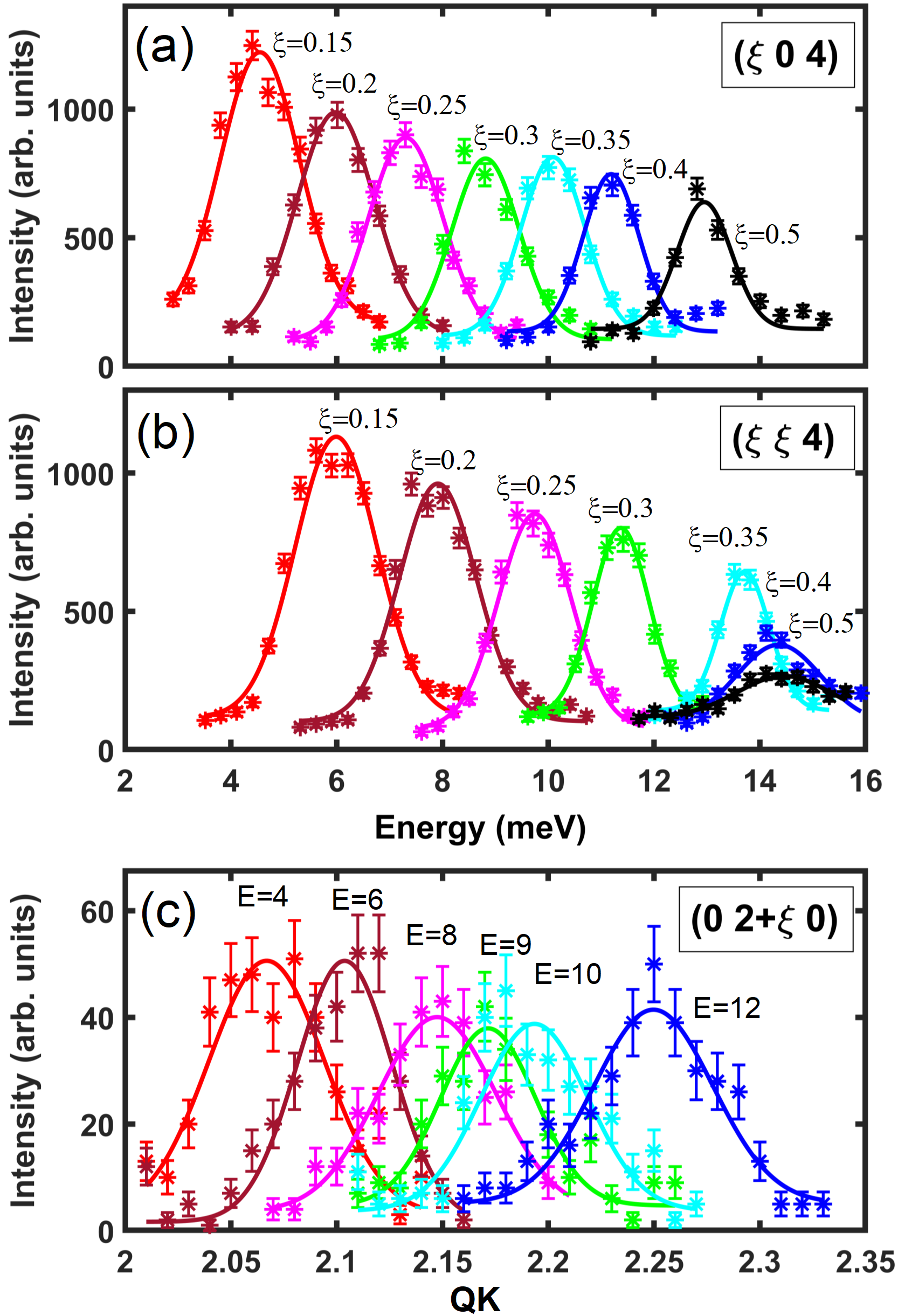}
		\caption{Measurements of acoustic $\Delta_4$ modes performed at 1.6\,K at the scattering vectors ($\xi$ 0 4) on IN8 (a), and of $\Sigma_4$ modes observed at ($\xi$ $\xi$ 4) on IN8 (b). The mode intensity diminishes with increasing $\xi$ due to the  factor of 1/$\omega$ in Eq. (1). Data were normalized to a monitor counting of 50,000, which corresponds to approximately 70 seconds counting time.  (c) INS constant-energy scans for the longitudinal $\Delta_1$ modes performed at (0\,2+$\xi$\,0) on 1T; these data are normalized to a  monitor counting of 30,000, which corresponds to about 35 seconds.}
		\label{acousticmodes}
	\end{center}
\end{figure}

Further experiments were performed in two cycles at the IN8 TAS at the ILL in Grenoble (France) \cite{data-in81,data-in82}. IN8 is a high-flux thermal neutron TAS designed to investigate a wide energy and momentum transfer range.
We used a Si(111) monochromator and a HOPG (002) analyzer, both in double-focusing mode. A HOPG filter in front of the analyzer suppressed higher-order contaminations. Scans were mostly performed with fixed final neutron momentum $k_f$\,=2.66\,\AA$^{-1}$ and in special cases with $k_f$\,=\,3.84\,\AA$^{-1}$. All reciprocal-space vectors are given in reduced lattice units with respect to the tetragonal lattice of $a$\,=\,$b$\,=\,3.765\,\AA \ and $c$\,=\,6.28\,\AA . 
After mounting the sample in the cryostat, Bragg reflections were used to orient the sample and to set the goniometers and diaphragms. Most INS experiments were performed at low temperature, $T$=1.6\,K, because the phonon broadening is typically smaller and the two-phonon scattering is reduced, which facilitates the study at high energies.
Besides, low-energy acoustic phonons exhibit smaller scattering intensity at low temperature due to the Bose factor, but these modes are not difficult to study. Typical scans to determine the phonon energies are shown in Fig.\,\ref{acousticmodes} and \ref{opticalmodes} for acoustic and optical modes, respectively. On both 1T and on IN8, the data were normalized by a monitor counter posed between the monochromator and the sample.

INS experiments can determine the phonon dispersion throughout the entire Brillouin zone but for complex materials with a large number $n$ of atoms in the primitive cell it is challenging to identify the character in terms of the phonon polarization for the individual modes. The interaction between the neutron and the ions is rather simple, described by the scattering length, and also the intensity of the coherent one-phonon scattering process can be easily calculated \cite{Shirane,Squires}:

\begin{equation}
		I \propto \frac{1}{\omega} \left(n(\omega)+\frac{1}{2}\pm\frac{1}{2}\right) \Bigg[\sum_d \frac{b_d}{\sqrt{m_d}} e^{(-W_d+i\textbf{Q.r}_d)} (\textbf{Q}\cdot\textbf{e}_d) \Bigg]^2
		\label{dynamical-struc-fac}
\end{equation}

where $\omega$ denotes the phonon angular frequency and $n(\omega)$ is the Bose factor; $\bf q$ is the phonon wave vector; $\bf{Q}=\bf{G}+\bf{q}$ is the scattering vector decomposed to a reciprocal lattice vector  $\bf G$ and the phonon propagation vector $\bf q$ in the first Brillouin zone. $e^{-W_d}$ is the Debye-Waller factor with atoms indexed by $d$, with mass $m_d$ and scattering length $b_d$. The $3n$-dimensional polarization vector  $\boldsymbol{e}=(\boldsymbol{e}_1,...,\boldsymbol{e}_{n})$ containing the $n$ polarization vectors of the individual atoms $\boldsymbol{e}_d$ enters the intensity through the scalar product $\bf{Q}\cdot\textbf{e}_d$, which indicates that intensity is maximal
for polarization components parallel or antiparallel to the scattering vector. 
The term in square brackets denotes the dynamical structure factor of the phonon mode at the specific scattering vector. 
The scattering intensity is proportional to $\omega^{-1}[n(\omega) +\frac{1}{2}\pm\frac{1}{2}]$, which implies that the lower energy modes yield stronger intensity (see Fig.\,\ref{acousticmodes}). One must distinguish phonon creation and annihilation processes that refer to the $+$ and $-$ of the $\pm$ sign, respectively.
At low temperatures ($\hbar\omega > k_B T$), the probability of phonon creation remains always finite while phonon absorption becomes suppressed by the Bose statistics. Therefore, such measurements must be performed in the neutron-energy-loss mode yielding the factor ($n(\omega)+1$). Individual phonons can be determined as a peak in the energy spectrum with respect to  $\bf{Q}$ values, thereby enabling one to map out the locus of a corresponding mode in the $E=\hbar\omega$ versus $\bf q$ plot.
The dynamical structure factor cannot easily be evaluated; just for acoustic phonons one immediately sees, that strong 
intensity requires the scattering vector to be parallel to the polarization vector and a large length $Q$ of the scattering vector close to a strong Bragg reflection.
With a lattice-dynamics model one can calculate the dynamical structure factors and by comparing measurements at 
various reciprocal lattice vectors $\bf G$ one can identify the polarization character of individual modes.

In the ideal case, the instrument resolution implies widths perpendicular to the scan direction that correspond to only little small impact on the phonon dispersion. This condition is reasonably well 
met for little dispersing optical phonons measured in a constant-$\bf Q$ scan, so that determining the peak positions by fitting against the intensity profile yields the correct points on a a dispersion area. However, for acoustic phonons with steep dispersion close to the Brillouin-zone center the resolution must be taken into account as it will be discussed in detail in the Appendix.  

\begin{table}[h]
	\centering
\begin{tabular}{|c|c|c|c|c|}
	\hline
	Temperature & 300 K & 220 K & 200 K & 90 K \\
	\hline
	a (\AA) & 3.7756(1) & 3.7735(1) & 3.7722(1) & 3.7703(1) \\
	c (\AA) & 6.3571(2) & 6.3384(2) & 6.3346(2) & 6.3190(2) \\
	$R_{\text{obs}}/R_{\text{all}}$ & 4.34/5.21 & 6.15/6.46 & 5.94/6.34 & 4.44/4.85 \\
	$wR_{\text{obs}}/wR_{\text{all}}$ & 7.01/7.03 & 7.55/7.55 & 7.67/7.68 & 6.89/6.90 \\
	GoF & 4.42/3.98 & 5.25/4.90 & 5.07/4.65 & 4.72/4.40 \\
	\hline
	Li 2c $(\frac{1}{4},\frac{1}{4},z)$ & 0.648(4) & 0.645(4) & 0.653(4) & 0.656(3) \\
	Li $U_{\text{iso}}$ & 0.023(4) & 0.017(4) & 0.017(4) & 0.012(3) \\
	As 2c $(\frac{1}{4},\frac{1}{4},z)$ & 0.2370(1) & 0.2368(2) & 0.2368(2) & 0.2365(2) \\
	As $U_{\text{11}}$ & 0.0092(3) & 0.0059(3) & 0.0080(3) & 0.0058(3) \\
	As $U_{\text{33}}$ & 0.0168(5) & 0.0112(5) & 0.0144(5) & 0.0121(4) \\
	Fe 2a $(\frac{3}{4},\frac{1}{4},0)$ & & & & \\
	Fe $U_{\text{11}}$ & 0.0087(4) & 0.0056(3) & 0.0078(4) & 0.0059(3) \\
	Fe $U_{\text{33}}$ & 0.0147(7) & 0.0100(6) & 0.0132(7) & 0.0112(6) \\
	\hline
\end{tabular}
	\caption{Crystal structure of LiFeAs in space group $P4/nmm$ determined from single-crystal X-ray diffraction experiments performed on a Bruker D8 diffractometer. Atomic displacement parameters $U$ are given in \AA$^2$.}
	\label{cryststructure}
\end{table}

\section{Crystal structure}

LiFeAs crystallizes in a tetragonal $P4/nmm$ structure, and there is no magnetic transition (see Table~\ref{cryststructure} and Fig.~1). 
The crystal structure was analyzed by X-ray diffraction experiments on a small (100x120x40 $\mu$m) crystal cut from the large
single-crystalline ingots that was mounted in a 150\,$\mu$m cryoloop. Cutting the material to yield a sample for X-ray single-crystal diffraction is delicate, and a suitable crystal could only be obtained in numerous attempts.
Measurements were performed on a Bruker D8 diffractometer equipped with micro-focus Mo source and with a $photon$ area  detector. The temperature was controlled with a nitrogen blower. 
At 90,
200, 220 and 300 K large data sets of Bragg reflection intensities
were collected with 10615(15716), 9038(14055), 11591(16585) and
9723(17656) observed(observed and weak) reflections, respectively.
These data were corrected for absorption and
merged into sets of 224(258), 219(258), 226(258) and 210(259) independent
reflections.
In the calculated precession maps we could not detect any evidence
for superstructure peaks breaking the $P4/nmm$ translation lattice.
Li$^+$ is hardly visible in X-ray experiments due to its low number of electrons. With the 90\,K data we verified the stoichiometry by refining the Li and As occupation numbers yielding 0.92(9) and 0.984(8), respectively. The deficiency of the Li site remains below its error and is not significant. Also
at the other temperatures, the refinement of the occupation did not indicate any deviation nor improved the fit. In the following analyses ideal occupation was assumed. Refinements of an extinction model yielded significant correction only for the (0\,0\,1) reflection; this reflection was excluded from the final refinement without further extinction correction.
Refinement of anisotropic atomic displacement parameters for As and Fe atoms 
improved the data description while for Li an isotropic value was assumed.
The results agree with previously published structural parameters \cite{Pitcher2008,Morozov2010,Brand2014}. 
Upon cooling from 300 to 90\,K we find only a small reduction of the As $z$ parameter and also the thermal expansion of the $a$ lattice parameter is small, $\Delta a/a$=0.0014. However the relative shrinking of the $c$ lattice constant is six times larger, $\Delta c/c$=0.006, which reflects the weak structural coupling between the layers. The atomic displacement parameters appear normal and shrink upon cooling. There is thus 
no indication for a structural phase transition in LiFeAs, while its crystal structure flattens upon cooling.

\section{Lattice-dynamics model}

LiFeAs has two formula units in the primitive cell and thus $n$\,=\,6 atoms. Therefore, there are 18 phonon modes at each propagation vector unless some modes are degenerate. 
The main challenge in quantitatively analyzing the phonon dispersion
of such a system relies in the proper identification of the neutron signals in terms of the phonon polarization vectors, which can be realized by Eq.~(1). Only by properly choosing the Brillouin zones of the measurements and by
comparing several measurements in different Brillouin zones with the calculations according to Eq.~(1) one may obtain 
a reliable picture of the phonon dispersion. Most importantly, the symmetry of the system, which separates the phonon modes according to irreducible representations, simplifies the identification of the modes. The distinction of
different representations is also important for visualizing the dispersion. Note, that branches belonging to different
representations can cross, but that two branches of the same representation interact and thereby avoid the crossing~\cite{Bruesch2012}. 

The 18 phonon branches are characterized by their polarization vectors, where symmetry restrictions cause degeneracy for the $\Lambda$ path and at Brillouin zone boundaries as well as at $\Gamma$. The $P4/nmm$ symmetry yields the irreducible representations for the $\Gamma$ modes that are illustrated in Fig.~\ref{pol-vectors}:

\begin{equation}
	\Gamma : 2A_{1g}+3A_{2u}+1B_{1g}+3E_{g}+3E_{u}.
\end{equation}

Here, $A$ and $B$ represent the non-degenerate symmetric, $g$, or anti-symmetric, $u$, vibrations, respectively, while $E$ represents the doubly degenerate modes. The $A$ modes denote vibrations along the $\bf c$ direction and have bond-stretching character with out-of-phase ($A_{1g}$ for Li and As) and in-phase polarization ($A_{2u}$ for Li, Fe and As). The $B_{1g}$ mode corresponds to the Fe vibration along the $\bf c$ direction and is called the buckling mode. 
$E_{g}$ and $E_{u}$ symbols refer to the symmetric (out of phase) or antisymmetric (in phase) modes of all atoms that
are polarized parallel to the layers. The acoustic modes correspond to the $u$ symmetries, because all atoms must 
move in phase.
The polarization schemes are presented in Table \ref{tablepol} for the $\Gamma$ modes and for the main symmetry directions, $\Delta$, $\Sigma$ and $\Lambda$. In our notation the irreducible representation
with the index 1 contains the longitudinal mode and those with indices 3 or 4 contain each one transverse acoustic mode. 
The in-plane polarized transverse acoustic modes always correspond to the index 3.
Note that the two sites for each ion type are related by the $n$ glide-mirror symmetry element. In the even $g$ modes these pairs possess opposite phases, while they are in-phase for the odd $u$ modes. Table II also gives the compatibility relations for the main symmetry directions.

\begin{figure}[h]
	\includegraphics[width=0.94\columnwidth]{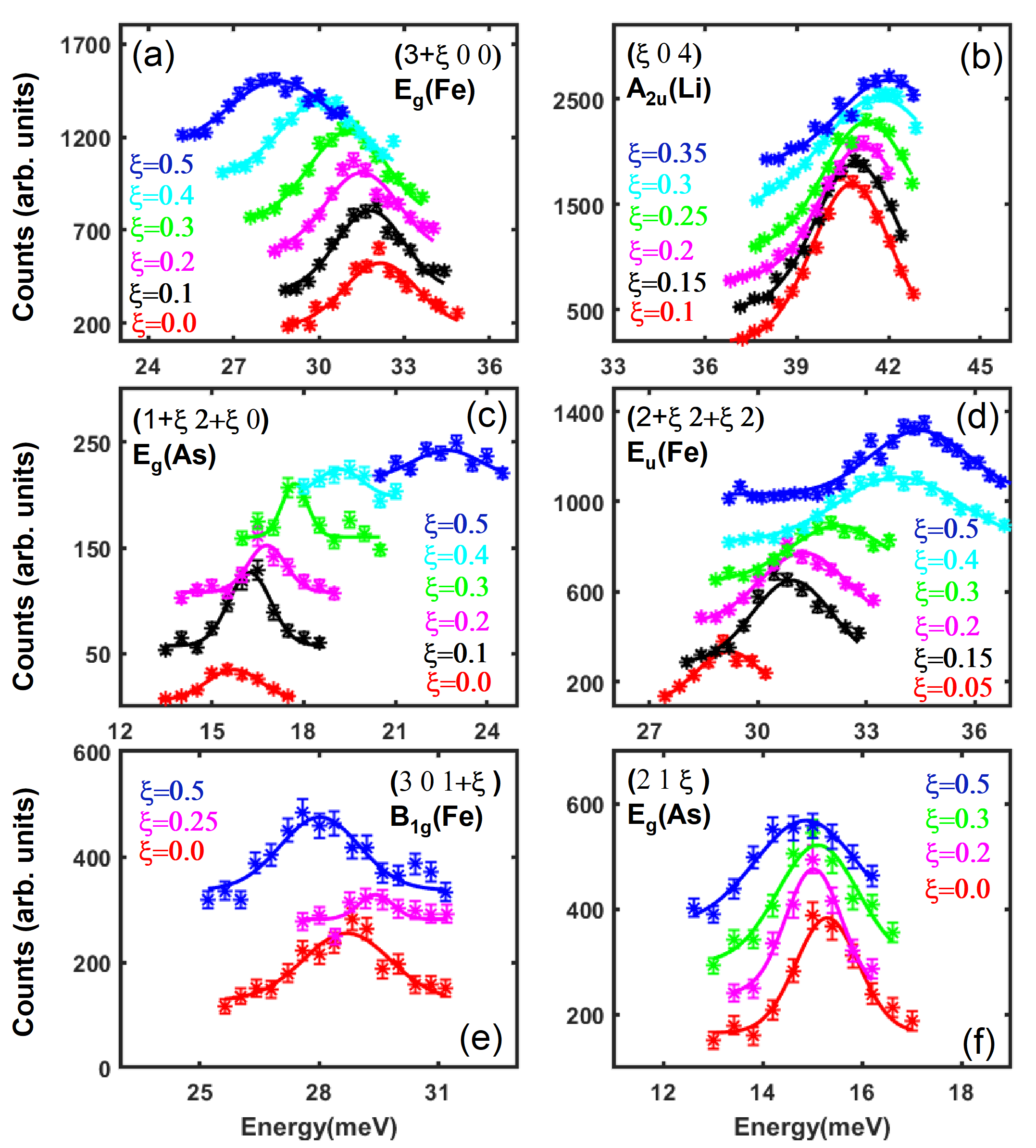}
	\caption{Measurements of optical modes with selected constant-$\bf{Q}$ scans for (a-b) q=($\xi$ 0 0); (c-d) ($\xi$ $\xi$ 0) and (e-f) (0 0 $\xi$). Data were normalized to a monitor counting of 50.000, which corresponds to approximately 80 to 90 seconds counting time.}
	\label{opticalmodes}
\end{figure}

\begin{figure*}[t!]
	%	\begin{center}
		\includegraphics[width=0.91\textwidth]{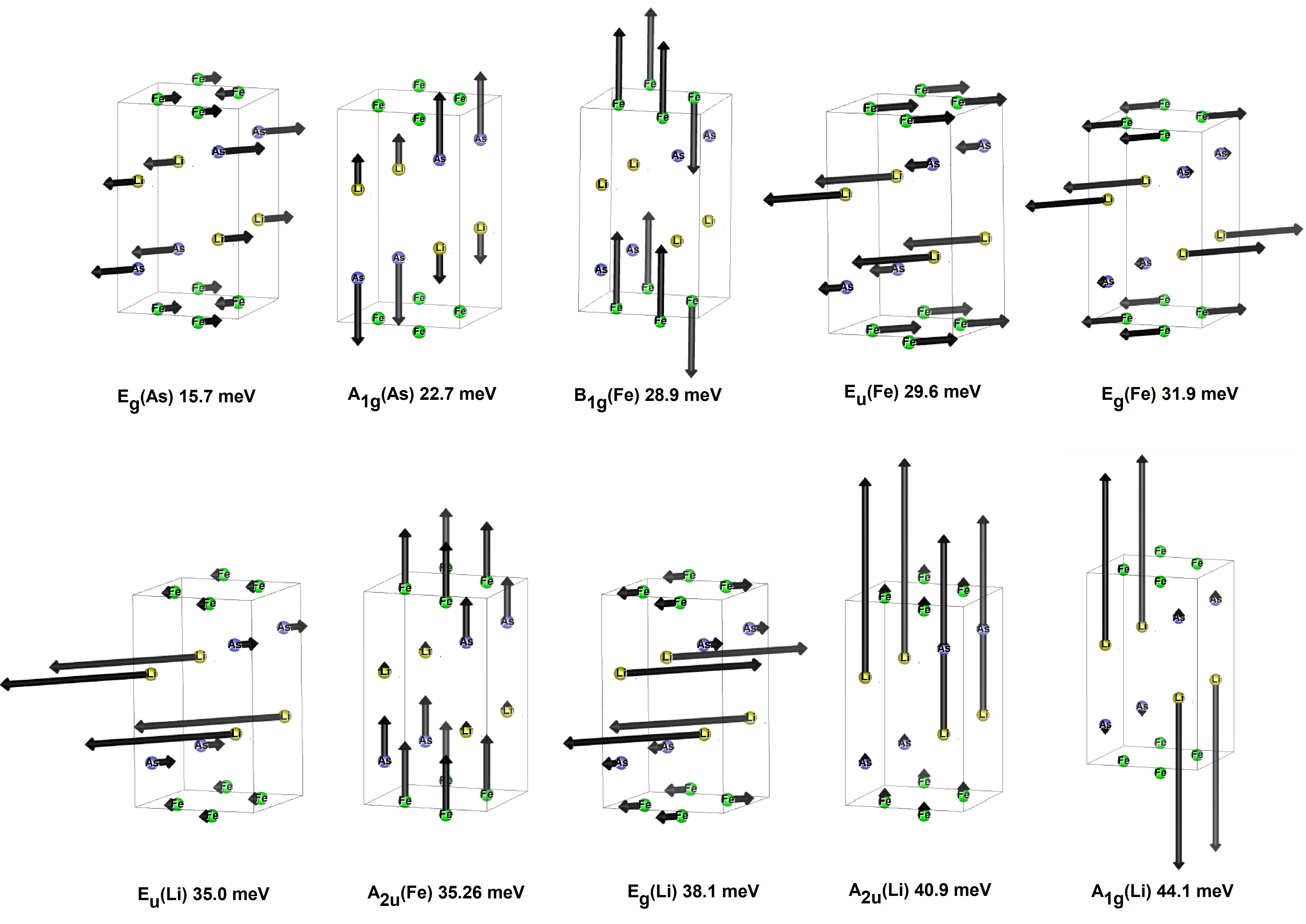}
		\caption{Phonon polarization patterns for all optical modes at the $\Gamma$ point of the Brillouin zone. The arrows indicate the amplitudes of the oscillation that contain the mass factor of the individual ions (Li yellow, Fe green, As blue). The label in brackets indicate the ion with the strongest contribution to the mode, and the numbers give the mode energies determined by the INS studies.}
		\label{pol-vectors}
		%	\end{center}
\end{figure*}

\begin{table*}[]
	\centering
	\begin{tabular}{|c| c c c c c c|}
		\hline
		& \textbf{Li1} & \textbf{Li2} & \textbf{Fe1} & \textbf{Fe2} & \textbf{As1} & \textbf{As2} \\
		& (0.25 0.25 0.6356) & (0.75 0.75 0.3644) & (0.75 0.25 0) & (0.25 0.75 0) & (0.25 0.25 0.2371) & (0.75 0.75 0.7629) \\
		\hline
		2A$_{1g}$ & 00A & 00-A & 000 & 000 & 00B & 00-B     \\
		%\hline
		3A$_{2u}$ & 00A & 00A & 00B & 00B & 00C & 00C     \\
		%\hline
		1B$_{1g}$ & 000 & 000 & 00A & 00-A & 000 & 000     \\
		%\hline
		3$\cdot$2E$_{g}$ & A00 & -A00 & B00 & -B00 & C00 & -C00    \\
		%\hline
		3$\cdot$2E$_{u}$ & A00 & A00 & B00 & B00 & C00 & C00    \\
		\hline
		6$\Delta_1$ & A0B & A0-B & C0D & C0-D & E0F & E0-F \\
		%\hline
		3$\Delta_2$ & 0A0 & 0-A0 & 0B0 & 0-B0 & 0C0 & 0-C0 \\
		%\hline
		3$\Delta_3$ & 0A0 & 0A0 & 0B0 & 0B0 & 0C0 & 0C0 \\
		%\hline
		6$\Delta_4$ & A0B & -A0B & C0D & -C0D & E0F & -E0F \\
		\hline
		5$\Sigma_1$ & AAB & AA-B & CC0 & CC0 & DDE & DD-E \\
		%\hline
		3$\Sigma_2$ & A-A0 & -AA0 & BB0 & -C-C0 & DDE & DD-E \\
		%\hline
		4$\Sigma_3$ & A-A0 & A-A0 & B-BC & B-B-C & D-D0 & D-D0 \\
		%\hline
		6$\Sigma_4$ & AAB & -A-AB & C-CD & -CCD & EEF & -E-EF \\
		\hline
		5$\Lambda_1$ & 00A & 00B & 00C & 00C & 00D & 00E \\
		%\hline
		1$\Lambda_2$ & 000 & 000 & 00A & 00-A & 000 & 000 \\
		%\hline
		6$\cdot$2$\Lambda_3$ & A-B0 & -CD0 & E-F0 & -GH0 & I-J0 & -KL0 \\
		\hline \hline
		& [100] & $\Delta_1$= 3E$_u$ + 2A$_{1g}$ +B$_{1g}$ & $\Delta_2$= 3E$_g$ & $\Delta_3$= 3E$_u$ & $\Delta_4$= 3E$_g$ + 3A$_{2u}$ & \\
		
		& [110] & $\Sigma_1$= 3E$_u$ + 2A$_{1g}$ & $\Sigma_2$= 3E$_g$ & $\Sigma_3$= 3E$_u$ + B$_{1g}$ & $\Sigma_4$= 3E$_g$ + 3A$_{2u}$ & \\
		
		& [001] & $\Lambda_1$= 3A$_{2u}$ + 2A$_{1g}$ & $\Lambda_2$= B$_{1g}$ & $\Lambda_3$= 3E$_u$ + 3E$_g$ & & \\
		\hline
	\end{tabular}
	\caption{Upper part: Polarization schemes according to the $P4/nmm$ crystal structure of LiFeAs for all $\Gamma$ modes and for the representations along the [100], [110], and [001] directions (labeled $\Delta$, $\Sigma$, and $\Lambda$, respectively); the first line gives the positions of the six atoms forming a primitive unit, the following lines show the displacements of these atoms for the respective irreducible representation. A letter at the $i$th position of an atom, signifies that this atom is moving along the $i$ direction, a second appearance of the same letter signifies that the second atom moves with the same amplitude (“-” denotes a phase shift) in the corresponding direction. Lower part: Compatibility relations of the irreducible representations are given for phonon wave vectors along the main-symmetry directions.}
	\label{tablepol}
\end{table*}

We used the program $GENAX$ \cite{genax} to model the lattice dynamics in LiFeAs. The harmonic approximation reduces the lattice-dynamics problem to the diagonalization of a $3n$$\times$$3n$ matrix for each allowed value of the phonon propagation vector $\bf q$  \cite{Bruesch2012}: $\omega^2{\bs e}={\bs \bar{D}{\bs e}}$ with $\bs e$ the 3$n$ dimensional phonon polarization vector and $\omega$ the phonon frequency.  The dynamical matrix ${\bs \bar{D}}$ is given by the sum over the force constants $\Phi_{\alpha,\beta}({\bs 0}d,{\bs l'}d')$ that relate a displacement of the atom $d'$ in the $\bs l'$ unit cell in $\beta$ direction to a force on the atom $d$ in the $\bs 0$ unit cell in $\alpha$ direction ($\alpha$,$\beta$\,=\,$x,y,z$):

\begin{equation}
	{\bar{\bs D}}_{\alpha,\beta}(d,d')=\frac{1}{(m_dm_{d'})^{1/2}}\sum_{{\bs l'}}{ \Phi_{\alpha,\beta}({\bs 0}d,{\bs l'}d')\exp{(i{\bs q \bs l'})}   }.
\end{equation}

The force constants for a pair of atoms $\Phi_{\alpha,\beta}({\bs 0}d,{\bs l'}d')$ form a 3$\times$3 matrix that one needs to determine. In the case of the two-ion potential energy depending only on the length of the distance, $\phi(r)$, there
are only two parameters, $F=\phi''(r)$ and $G=\phi'(r)/r$, determining the force-constant matrix for a pair, which sometimes are called longitudinal and transversal constants:

\begin{equation}
\Phi_{\alpha,\beta}({\bs 0}d,{\bs l'}d')=-\frac{r_\alpha r_\beta}{r^2}(F-G)-\delta_{\alpha \beta}G,
\end{equation}

where $r_\alpha$ and $r_\beta$ are the coefficients of the distance vector $\bf r$  between the two ions.
In ionic compounds the force constants can be deduced from empirical potentials, which can simulate
the long-ranging Coulomb potentials either directly or by introducing mass-less electron shells \cite{Bruesch2012}. Such models 
were successful to describe the lattice dynamics in superconducting cuprates \cite{Chaplot1995} and Sr$_2$RuO$_4$ \cite{Braden2007}.
For the lattice dynamics of \lfa we directly refined 24 force constants to
describe the interaction between ions for distances up to 4.3 Å. The first approach used the phonon frequencies determined in Raman experiments \cite{Um2012} and the acoustic modes obtained from our INS experiments. 
The characters of optical modes at high-symmetry paths
were then identified by model calculations of the INS intensity following the symmetry analysis in Table \ref{tablepol}. 
With the $GENAX$ program we refined the force constants with modes that exhibit a strong and unique dynamical structure factor at the studied $\bf Q$,$E$ range. 
The experimental phonon dispersion is compared in Fig.\,\ref{phonondispersion} with the DFT results.
The maximum deviation obtained between the calculated and experimental energies amounts to $\sim$3\,meV.
The comparison between experimental energies and the force-constant model calculations is given in the APPENDIX.

\section{DFT calculations}

The lattice dynamics of LiFeAs were further investigated using first principles DFT based techniques to elucidate the character of the different branches. Phonon properties were calculated within the linear response or density-functional perturbation theory (DFPT) implemented in the mixed-basis pseudopotential method \cite{Meyer,Heid1999}. The electron-ion interaction is described by norm-conserving pseudopotentials, which were constructed following the description of Vanderbilt~\cite{Vanderbilt1985}. Semicore states Fe-$3s$ and Fe-$3p$ were included in the valence space. In the mixed-basis approach, valence states are expanded in a combination of plane waves and local functions at atomic sites, which allows an efficient description of more localized components of the valence states. Here, plane waves with a cut-off for the kinetic energy of 24\,Ry and local functions of $s$, $p$, $d$ type for Fe were employed.
For the exchange-correlation functional the general-gradient approximation (GGA) in the Perdew-Burke-Ernzerhof form~\cite{PBE} was used. Brillouin zone integrations were performed with a 16x16x8 $\bf k$-point grid combined with a Gaussian smearing of 0.1\,eV.

For the phonon calculations, structural parameters after Morozov et al. \cite{Morozov2010} were adopted. It is known from other iron-pnictide families, that for structural optimization with standard correlation functionals (LDA or GGA), the As-$z$ parameter, which determines the distance of the As atom to the Fe plane, deviates significantly from the experimental value. This behavior is found for LiFeAs, too. Within GGA, the As-$z$ value relaxes to 0.2197, as compared to the experimental value of 0.2363. This corresponds to a rather large error of about 0.1\,\AA ~ in the As position. For the phonon calculations, we therefore used the experimental values also for the positional parameters. The effects of structure relaxation on the phonon dispersion are discussed in the Appendix.

To obtain phonon properties, dynamical matrices were calculated by DFPT on a 8x8x4 momentum grid, and then determined for arbitrary points in the Brillouin zone by standard Fourier-interpolation techniques, providing both phonon frequencies and eigenvectors. Fig.~\ref{phonondispersion} compares the experimental results with the DFPT calculations. The good agreement with most measured phonon branches suggests a high accuracy of the DFT model for the lattice dynamics of LiFeAs. The total and partial phonon density of states shown in Fig.~\ref{pdos} was obtained within the same model by sampling the tetragonal Brillouin zone on a 40x40x20 grid using the tetrahedron method.

To assess the strength of the electron-phonon coupling, the DFPT approach was applied to calculate the electron-phonon coupling matrix elements on the same 8x8x4 phonon mesh. To achieve proper convergence, a denser $\bf k$-point mesh of 32x32x16 was employed. 
Summing the partial contribution to the electron-phonon coupling from each mode
over the 18 phonon branches and averaging on the Brillouin zone yields the total electron-phonon coupling constant of 0.19, which agrees with \cite{Boeri2008} and indicates a very small electron-phonon coupling in LiFeAs.

\section{Phonon dispersion in LiFeAs and its temperature dependence}

\subsection{Comparison of the INS and DFT phonon dispersion }

\begin{figure*}[tbp]
	\begin{center}
		\hspace{-.75cm}
		\includegraphics[width=0.94\textwidth]{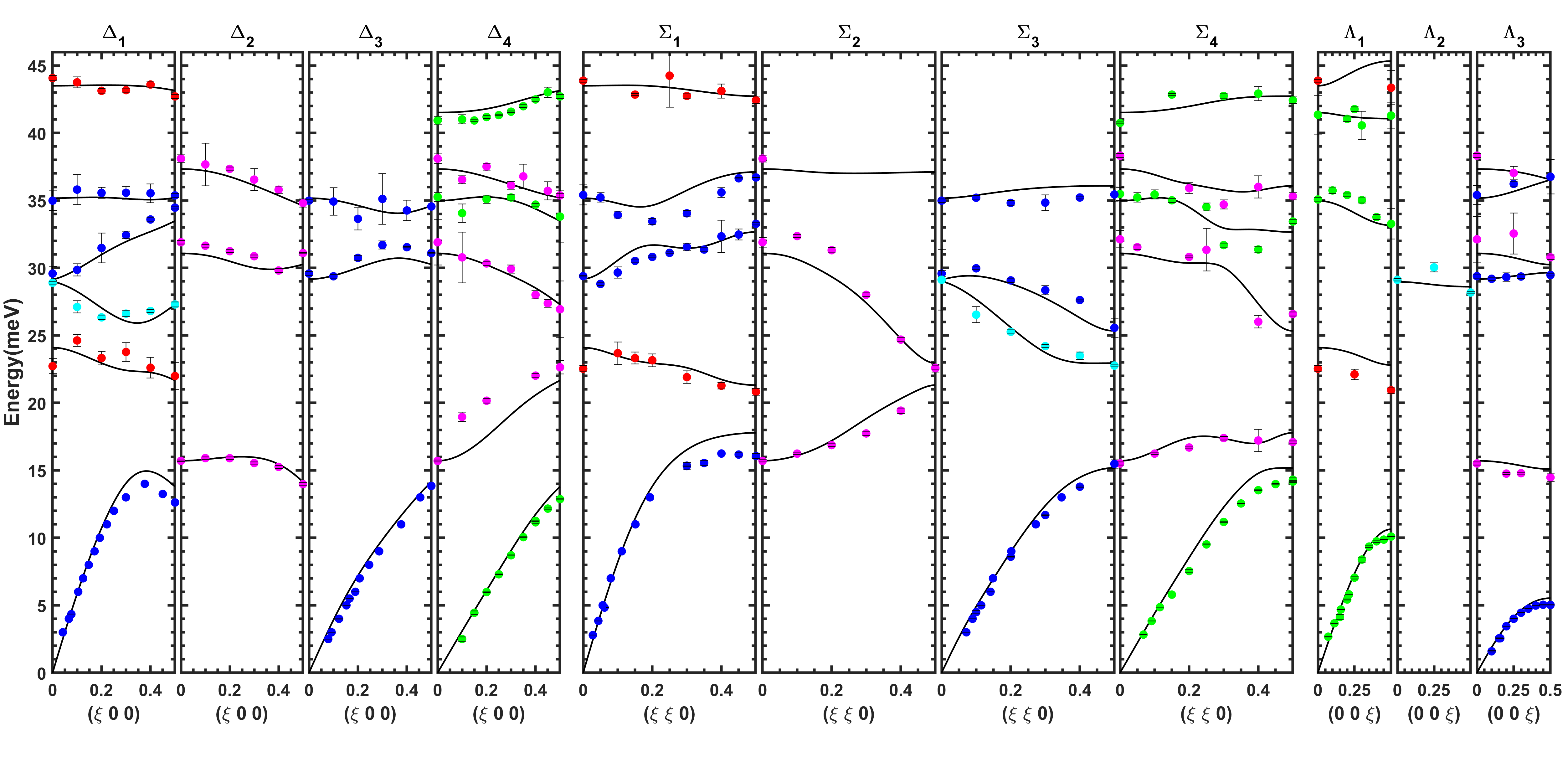}
		\caption{Phonon dispersion in LiFeAs. Symbols denote the energy values determined by INS and the black solid lines give the results of the DFT calculations. The dispersion is presented along the high symmetry directions $\Delta$ along [$\xi$ 0 0]; $\Sigma$ along [$\xi$ $\xi$ 0] and $\Lambda$ along [0 0 $\xi$] and the color indicates the symmetry of the $\Gamma$ mode at which the branch starts: red-$A_{1g}$; green-$A_{2u}$; cyan-$B_{1g}$; magenta-$E_g$; blue-$E_u$}
		\label{phonondispersion}
	\end{center}
\end{figure*}

Figure~\ref{acousticmodes} illustrates how efficiently acoustic phonons can be studied even at low temperature due to their high signal strength stemming from the $\frac{1}{\omega}$ term in Eq.~(1). Strong acoustic phonon signals appear at strong
Bragg peaks  $\bf G$ with $\bf Q$\,=\,${\bf G}+{\bf q}$ determining the transversal or longitudinal polarization of the modes due to the factor ${\bf Q}\cdot{\bs e_d}$ in Eq.~(1). To measure a particular optical phonon it is crucial to find a $\bf G$ reciprocal lattice vector, for which this mode obtains a strong dynamical
structure factor while all other modes with close-lying energies vanish. On a TAS instrument, the experiment on optical phonons must be guided by predictions from a lattice-dynamics model, that can be successively refined. Fig.\,\ref{opticalmodes} presents several measurements on various optical modes that yield single peaks and thus a clear identification of the character.

Figure~\ref{phonondispersion} shows the measured phonon dispersion compared to the DFT calculations. The data have been plotted with error bars for the respective mode at the $\bf q$ positions with the maximum error being around 2.2\,meV. 
Overall there is good agreement between the dispersion obtained in the INS experiments and the DFT calculations.
The deviation between the DFT model and the mode energies observed by INS was calculated in terms of relative deviations, with the maximum difference amounting to 5.2\% for the E$_g$(As) mode at $\Delta_4$. 

A comparison of our INS data with the calculated dispersion from Ref. \cite{jishi,huang}  shows significant differences. In Ref.~\cite{jishi}, there is a pronounced instability in the longitudinal and transverse acoustic modes at the Z point, (0 0 0.5), where the phonon frequencies almost fall back to zero. This behavior is excluded by our INS data that does not indicate any zone-boundary instability and it is also not seen in the calculation in Ref. \cite{huang} nor in our DFT analysis. However, the latter calculation in Ref. \cite{huang} finds all phonon frequencies at energies below ~38\,meV in contradiction with our measurements and also in contradiction with the Raman determination of the $A_g$ mode energy \cite{Um2012}, see below. Thus, only our DFT calculations yield satisfactory agreement.

In centrosymmetric LiFeAs, zone-center phonons can be separated in even, $g$, and odd, $u$, modes that keep and
break inversion symmetry, respectively, and that are in principle Raman and infrared active, but the metallic
properties of LiFeAs exclude precise infrared studies on phonons.
We find the two optical $A_{2u}$ modes at 40.9\,meV and 35.3\,meV that exhibit essential Li and Fe polarization, respectively, compare Fig.~\ref{pol-vectors}. The two optical $E_u$  modes at 35.0\,meV and 29.6\,meV also exhibit mainly Li and Fe polarization, respectively. In the following we indicate the main ion contributing to the vibration by the label in brackets, see also Fig.~\ref{pol-vectors}. The $u$ modes are Raman inactive and have not been reported previously,  but for six of the $g$ modes we may compare the INS and the Raman results.
The $A_{1g}$ energies were determined in the INS experiments to 44.1\,meV (Li) and 22.7\,meV (As) while the 
Raman study observes these modes at 42.2\,meV (Li) and 23.1\,meV (As), respectively, yielding good agreement. Also the neutron result for the $B_{1g}$ energy at 28.9\,meV and the Raman result at 29.7\,meV (Fe) agree well. The two higher $E_g$ modes were determined at 38.1\,meV (Li) and 31.9\,meV (Fe) by INS compared to the Raman values of 38.5 and 36.1\,meV. The Raman result for the  $E_g$ (Fe) mode is clearly incompatible with our INS data questioning the proper interpretation of the very weak Raman signal, while the DFT calculation yields an energy close to the INS result. The Raman experiment also fails to observe the lowest $E_g$ energy that our INS scans at $\bf Q$=(1 2 0) unambiguously determine at 15.7\,meV at low temperature. 

\begin{figure}[h]
	\begin{center}
		% \hspace{-0.6cm}
		\includegraphics[width=0.84\columnwidth]{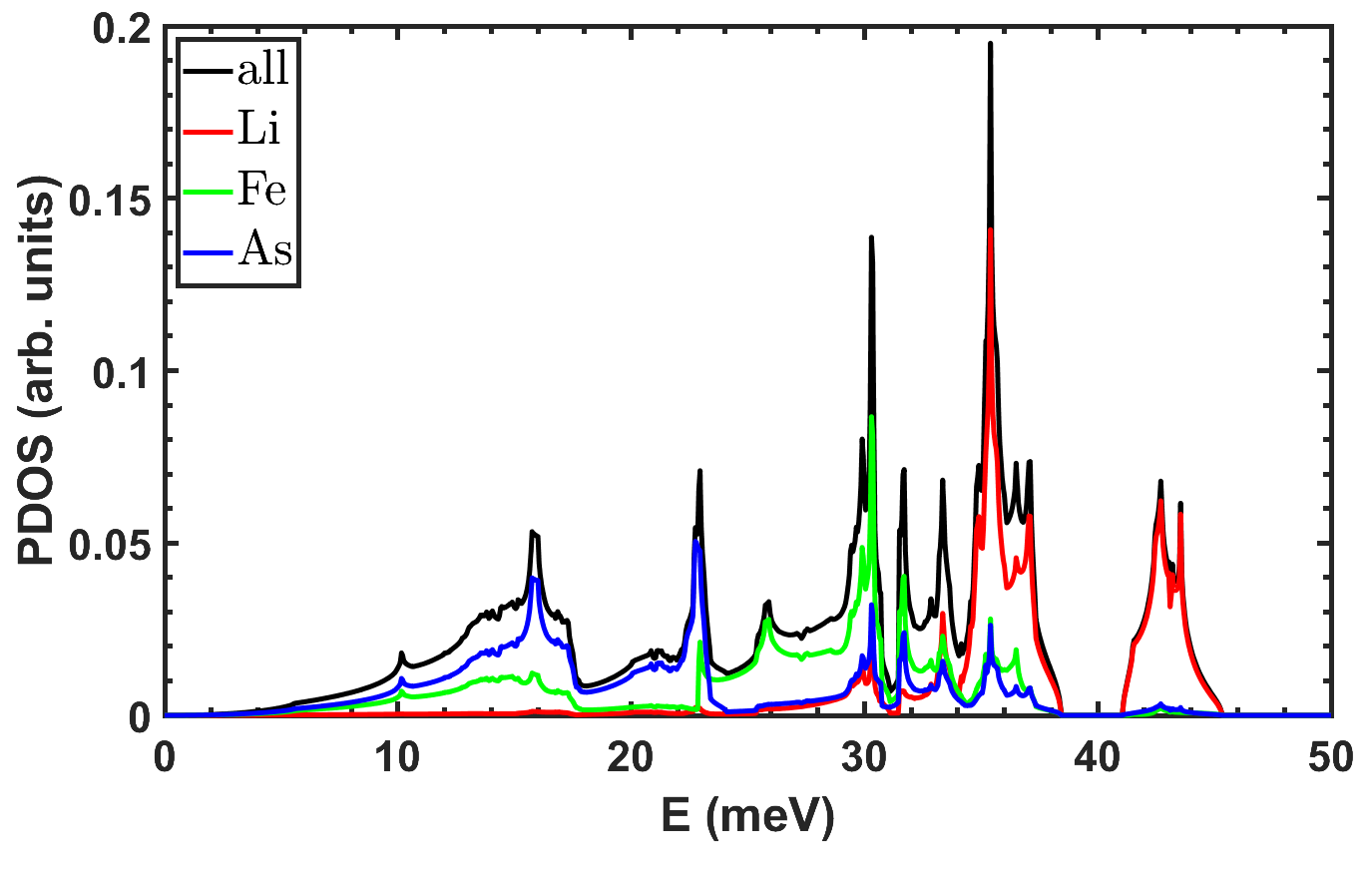}
		\caption{Phonon density of states in LiFeAs as calculated by DFPT. The black line
			gives the total density of states while red green and blue lines denote the partial density of states for Li, Fe and As, respectively.}
		\label{pdos}
	\end{center}
\end{figure}

In order to discuss the detailed aspects of the phonon dispersion it is helpful to inspect the phonon density of states (PDOS) calculated by DFT, see Fig.\,\ref{pdos}. We show the total PDOS as well as the partial PDOSs for the three different atoms Li, Fe and As. The highest branches contain the $A_{1g}$\,(Li)
mode and are separated from the rest of the dispersion. As indicated in the PDOS in Fig.\,\ref{pdos} and in the polarization pattern in Fig.\,\ref{pol-vectors} these modes correspond to almost pure Li vibrations perpendicular to the planes.
These branches at 42 to 44\,meV exhibit little dispersion resulting in a narrow band in the PDOS. For all lower branches there is no
clear separation visible as gaps in the PDOS. As it is expected from the large mass differences, further Li contributions
are found in the high energy mixed branches between 25 and 38\,meV, while even for the acoustic branches the Li oscillation is rapidly suppressed with increasing energy. The decoupling of the Li dynamics also corresponds to the structural arrangement with Li ions intercalating the FeAs layers. In contrast the mass ratio between Fe and As results in a less pronounced separation of modes, also because Fe and As form strongly connected structural units. Nevertheless one recognizes the change of dominating As contributions towards dominating Fe oscillations around 23\,meV in the PDOS.

The $A_{1g}$ (As) mode at 22.7\,meV involves the vertical even oscillation of the As and thus the modulation of the thickness of the FeAs layer, which was found to be decisive for the superconducting $T_c$~\cite{Lee2008}. A similar mode exists also in cuprates and ruthenates and the branch starting from this mode in the vertical direction ends at a peculiar mode, where the thickness in neighboring layers alternates. This zone-boundary mode can couple to inter-layer charge transfer and it is rather unusual in these latter compounds~\cite{Braden2007}.  However, LiFeAs does not show any indication for similar anomalous behavior.

At the $B_{1g}$(Fe) buckling mode at 28.9\,meV $\Delta_1$ and $\Sigma_3$ branches begin that soften in the Brillouin zone. For the $\Delta$ direction this effect even seems to cause interaction with the two lower branches resulting in a reduced frequency of the zone-boundary acoustic phonon. While this indicates some structural instability, the system is still far from such a structural phase transition. This aspect is well described by the DFT calculations as well as by the force constant model.

The $E_g$ modes involve even antiphase vibrations parallel to the FeAs layers and the two modes at lower energies exhibit a mixed character with Fe and As displacements. The  $\Sigma_2$ and $\Sigma_4$ branches starting at the $E_g$ mode at 31.9\,meV considerably soften towards the zone boundary yielding anticrossing behavior, see Fig.\,\ref{phonondispersion}. Also this softening is well reproduced by the DFT and by the lattice-dynamics calculations,see Fig.\,\ref{phonondispersion} and APPENDIX. Furthermore, it is not obvious to understand strong electron-phonon coupling for these modes.

The $u$ modes are infrared active and would imply a splitting between the longitudinal and transversal $\Gamma$ modes in an insulator. In metallic LiFeAs the long-range Coulomb potentials are screened so that this splitting must vanish at $\Gamma$. However, the screening is less effective at shorter distances corresponding to finite propagation vector, which results in an upwards dispersion of such polar-modes branches in simple metals~\cite{Braden2005}. The recovery of the longitudinal energy enhancement can indeed be seen for the $\Delta_1$ and $\Sigma_1$ branches starting at these $E_u$ modes, which indicates that the screening is far from being perfect in LiFeAs. For these modes the agreement between experiment and DFT energies is worse compared to the overall agreement and can point to some special coupling. In particular the end modes of acoustic $\Delta_1$ and $\Sigma_1$ branches exhibit energies considerably below the DFT calculations. Only for the $\Delta_1$ direction this behavior can be associated with the softening of the $B_{1g}$ branch.

Two ARPES measurements deduced different characteristic energy ranges for electron boson coupling.
Kordyuk et al.\,\cite{Kordyuk2011} report 15, 30 and 44\,meV and Li et al.\,\cite{Li2020} identify 20, 34, and 55\,meV.
The 55\,meV cannot be attributed to one-phonon scattering, while 44\,meV corresponds roughly to the $A_{1g}$ (Li) branch.
In the energy range 30 to 34\,meV many modes exist rendering it difficult to attribute the ARPES effects to a single phonon pattern.
However, 15\,meV corresponds to the end points of acoustic $\Delta_1$ and $\Sigma_1$ branches that exhibit also some
discrepancy with the DFT calculations.  Reference \citep{Kordyuk2011} elucidates the presence of van-Hove singularities enhancing the electronic density of states at the $\bf M$=(0.5\,0.5\,0) point.
Longitudinal phonon modes couple to charge modulations and can thus exhibit
strong electron-phonon coupling. For example the longitudinal bond-stretching modes in various perovskite compounds couple to charge-density modulations and typically exhibit an anomalous phonon dispersion \cite{Weber1987,Braden2005,Braden2002a}.

\begin{figure}[h]
	\begin{center}
		% \hspace{-0.6cm}
		\includegraphics[width=0.94\columnwidth]{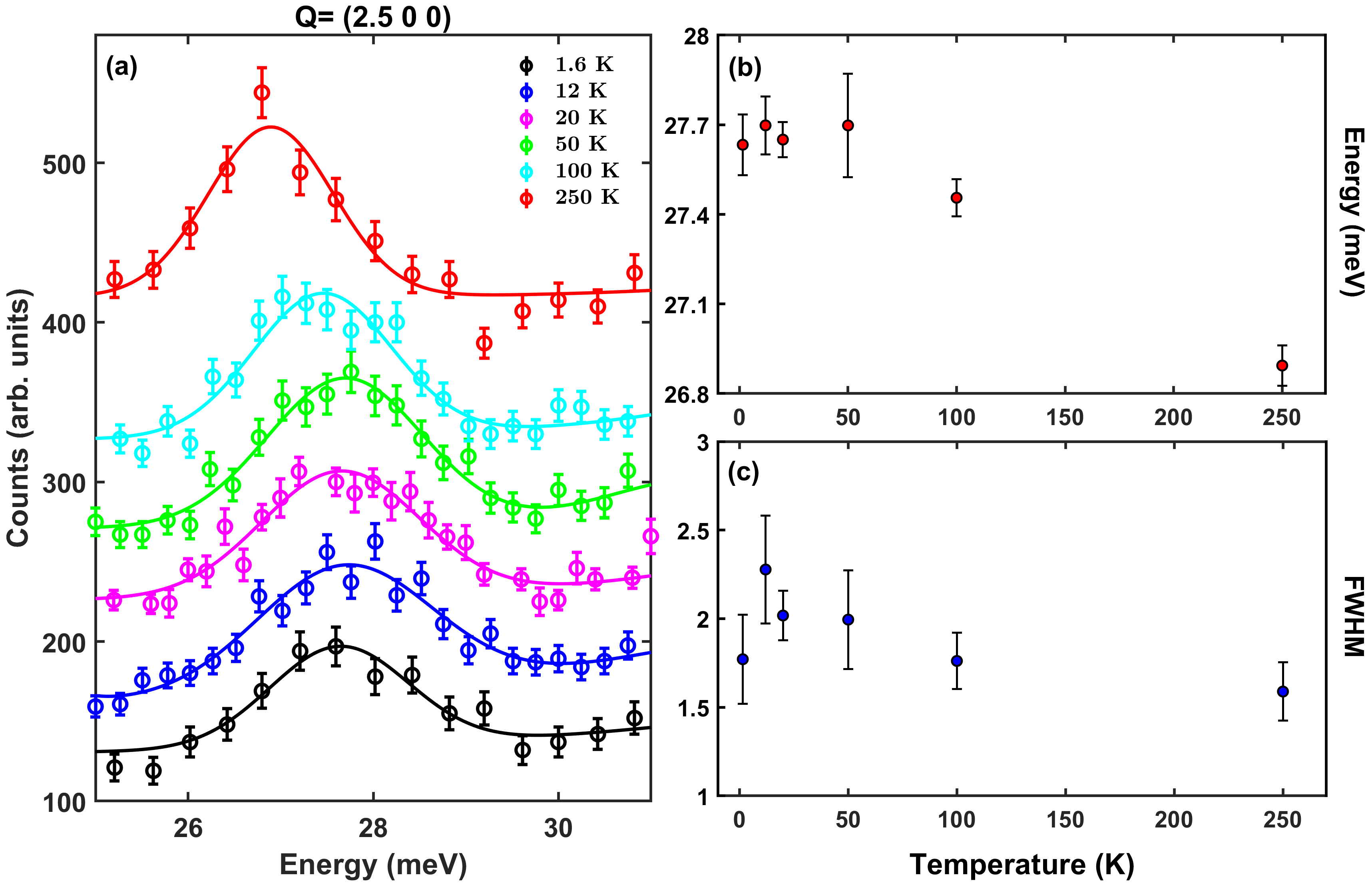}
		\caption{Temperature dependent scans for the $\Delta_1$ zone boundary mode at $\bf Q$\,=\,(2.5 0 0) that connects with the $B_{1g}$ mode (a). The peak energies (b) and their corresponding FWHMs (c) are plotted as a function of temperature.}
		\label{B1g_2p500}
	\end{center}
\end{figure}

\begin{figure}[h]
	\begin{center}
		% \hspace{-0.6cm}
		\includegraphics[width=0.94\columnwidth]{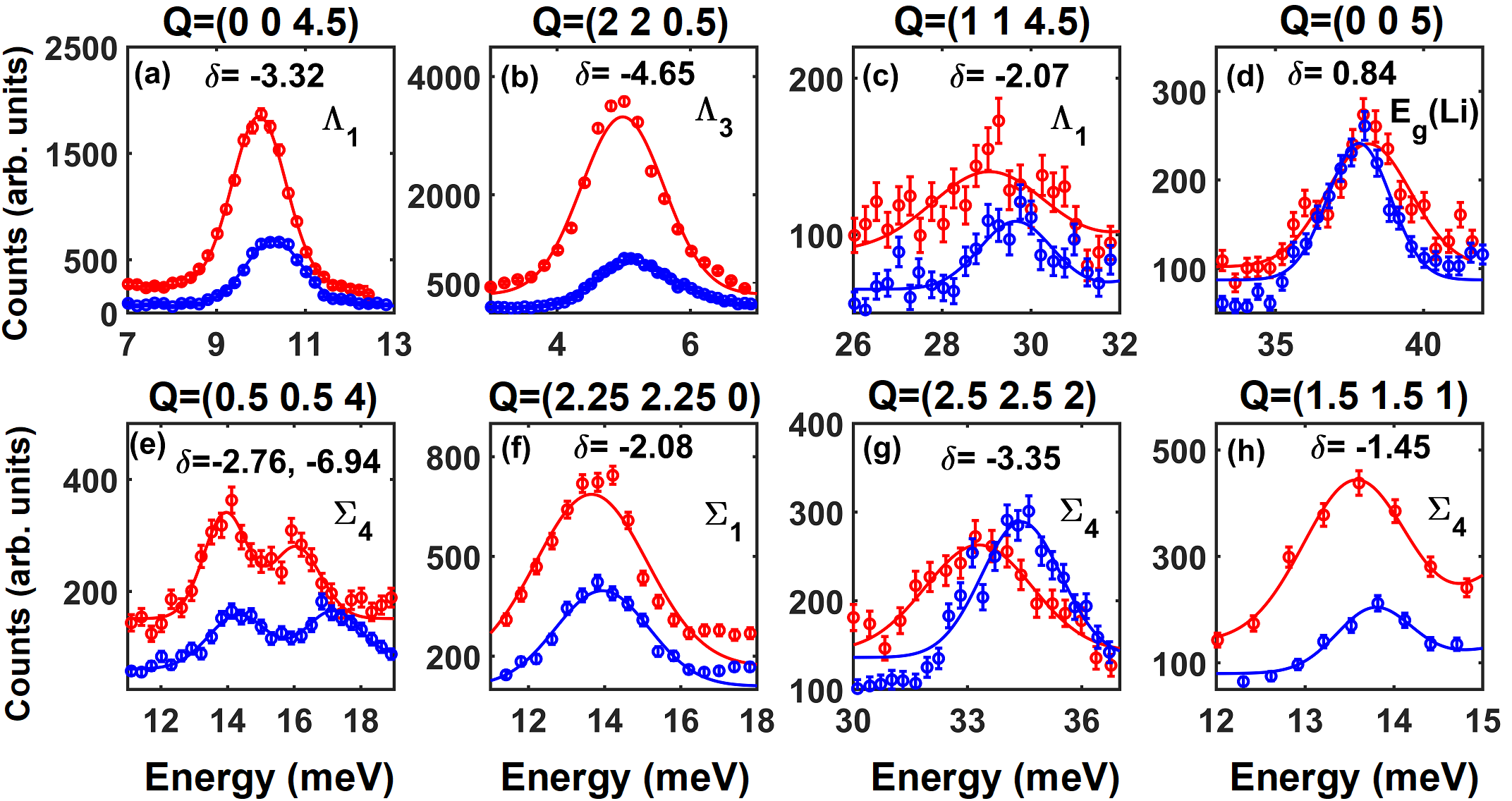}
		\caption{Temperature dependent scans for selected modes (a-h). $\delta$ denotes the relative peak shift for 1.6\,K (blue) and 290\,K (red) scans in percent with a negative value corresponding to low-temperature hardening of phonons.}
		\label{temp_dep_1}
	\end{center}
\end{figure}

\begin{figure}[h]
	\begin{center}
		% \hspace{-0.6cm}
		\includegraphics[width=0.94\columnwidth]{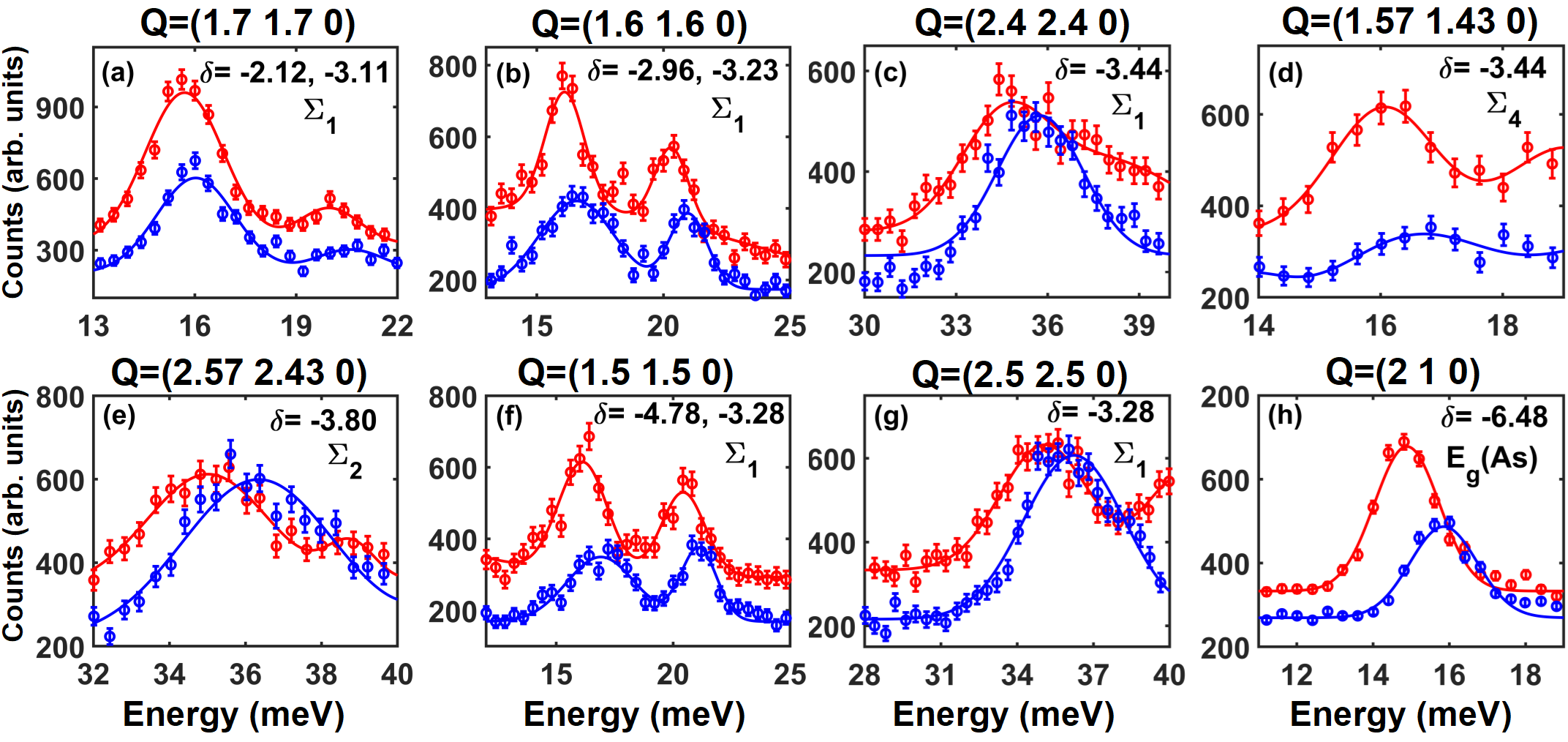}
		\caption{Temperature dependent scans for few selected modes (a-h). $\delta$ denotes the relative peak shift for the 1.6\,K (blue) and 290\,K (red) scans as in Fig.~\ref{temp_dep_1}.}
		\label{temp_dep_2}
	\end{center}
\end{figure}

\subsection{Temperature dependence of selected modes}

Raman experiments on LiFeAs report the temperature dependencies of five even zone-center modes~\cite{Um2012}. Upon cooling from room temperature to 5\,K, the $A_{1g}$, $B_{1g}$ and the highest $E_{g}$ modes exhibit a relative hardening by 2 to 3\% and become much sharper at low temperature. Both effects can be attributed to the normal coupling of phonon energies to the thermal expansion and 
to reduced phonon-phonon scattering effects on the line-widths. The Raman signal interpreted as the middle $E_{g}$ mode \cite{Um2012}, which disagrees with the INS analysis as discussed above, shows much smaller hardening and is always sharper than the other modes. The Raman experiments do not find any anomalous temperature dependence and thus confirm the absence of a structural phase transition~\cite{Um2012}.

We have studied the temperature dependencies of several selected phonon modes via constant-$\bf Q$ scans, see Fig.s ~\ref{B1g_2p500}, \ref{temp_dep_1} \ref{temp_dep_2}. First we searched for an impact of the superconducting transition on the
energy or full width at half maximum (FWHM) of the $\Delta_1$ modes at ($\xi$\,0\,4) with $\xi$\,=\,0.2, 0.25 and 0.5 that yield strong INS 
signals, see Fig.\,\ref{acousticmodes}. The data at 1.6 and 20\,K coincide within the errors in spite of the good statistics. Also the $A_{2u}$ mode at 36\,meV measured at (0 0 6), the lowest three $\Delta_1$ zone boundary modes measured at (2.5 0 0) and (2.5 0 2), the 38\,meV $\Delta_4$ mode at (0.3 0 5) and the 36\,meV $\Lambda_3$ mode at (3 0 1.5) do not exhibit any significant shift occurring at low temperature that would be induced by the opening of the superconducting gap in LiFeAs. 

However, we do observe significant changes when comparing room-temperature and low-temperature data in agreement with the Raman studies~\cite{Um2012}. The $c$ lattice parameter considerably shrinks upon cooling while
the in-plane thermal expansion is small resulting in a pronounced flattening of the crystal structure, see Table I.
Fig.\,\ref{B1g_2p500} shows the exemplary temperature dependence of the  $\Delta_1$ zone boundary mode near 27\,meV; the $\sim$3\% hardening follows the thermal expansion and can be attributed to a normal Grüneisen parameter.
In Fig.\,\ref{temp_dep_1} we compare room and low-temperature scans obtained with the first set of experiments on IN8. We calculate the relative change in the mode energy $\delta=\frac{E(290\,{\textrm K)}-E(1.6\,{\textrm K})}{E(1.6\,{\textrm K})}$ in \%; a negative number indicates hardening upon cooling. 
Albeit the $\delta$ values vary, the absolute range of energy shifts stay within the expectation for an anharmonic effect induced by thermal expansion whereas the variation of $\delta$ as such originates in the strong
anisotropy of the lattice contraction.
The lowest $E_{g}$(As) mode measured at (2 1 0)  hardens by 6.5\,\%, see Fig.\,\ref{temp_dep_2}(h), which is remarkable but most likely still a purely structural effect.

With the second set of experiments on IN8 we studied the temperature dependence of the $\Sigma_1$ modes close to the typical nesting vector (0.5 0.5 0) of FeAs-based materials, near which incommensurate magnetic excitations emerge in \lfa\,\cite{Qureshi2012}, see Fig.\,\ref{temp_dep_2}. Fig.\,\ref{temp_dep_2} also presents data for the $\bf{q}_{inc}$ position of the incommensurate magnetic excitations that is not lying on a high symmetry path. 
All these modes show significant hardening that is comparable to that of the other studied modes and the effects at the 
incommensurate propagation vector $\bf{q}_{inc}$ are not enhanced.

\begin{figure}[h]
	\begin{center}
		% \hspace{-0.6cm}
		\includegraphics[width=0.94\columnwidth]{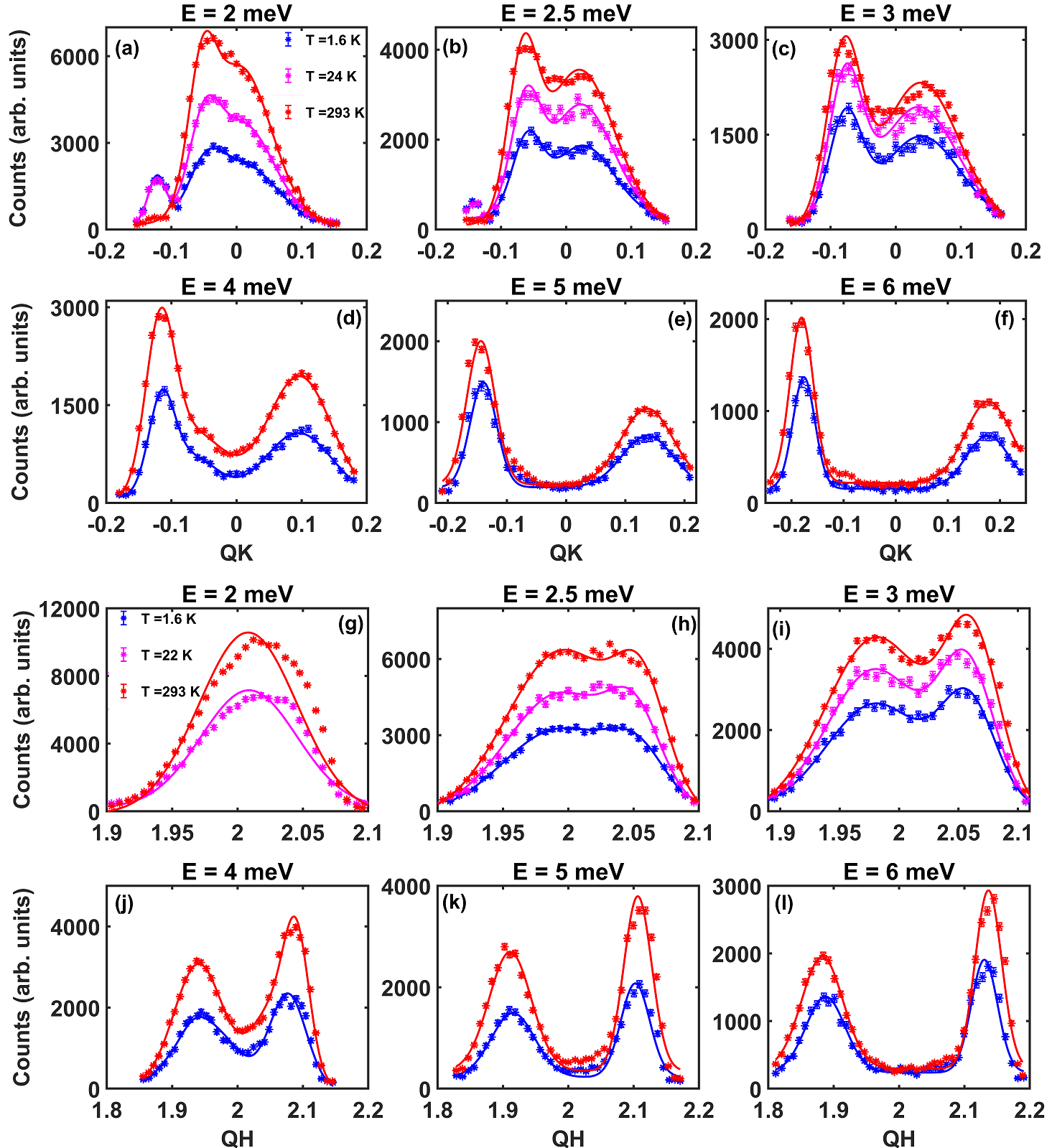}
		\caption{Constant-energy scans to investigate the in-plane polarized transversal acoustic modes of $\Delta_3$ and $\Sigma_3$ symmetry at scattering vectors (2 $\xi$ 0) (a-f) and at (2+$\xi$ 2-$\xi$ 0) (g-l).  Some extra contributions arise at low energy near the center because the finite resolution function also integrates other contributions. All low-temperature data were normalized to a monitor of 50.000 but the 293\,K data were normalized to 10.000 (a-c,j-k), 15.000(d,e,j,k) and 20.000 (f,l).}
		\label{const_energy}
	\end{center}
\end{figure}

\begin{figure}[h]
	\begin{center}
		% \hspace{-0.6cm}
		\includegraphics[width=0.94\columnwidth]{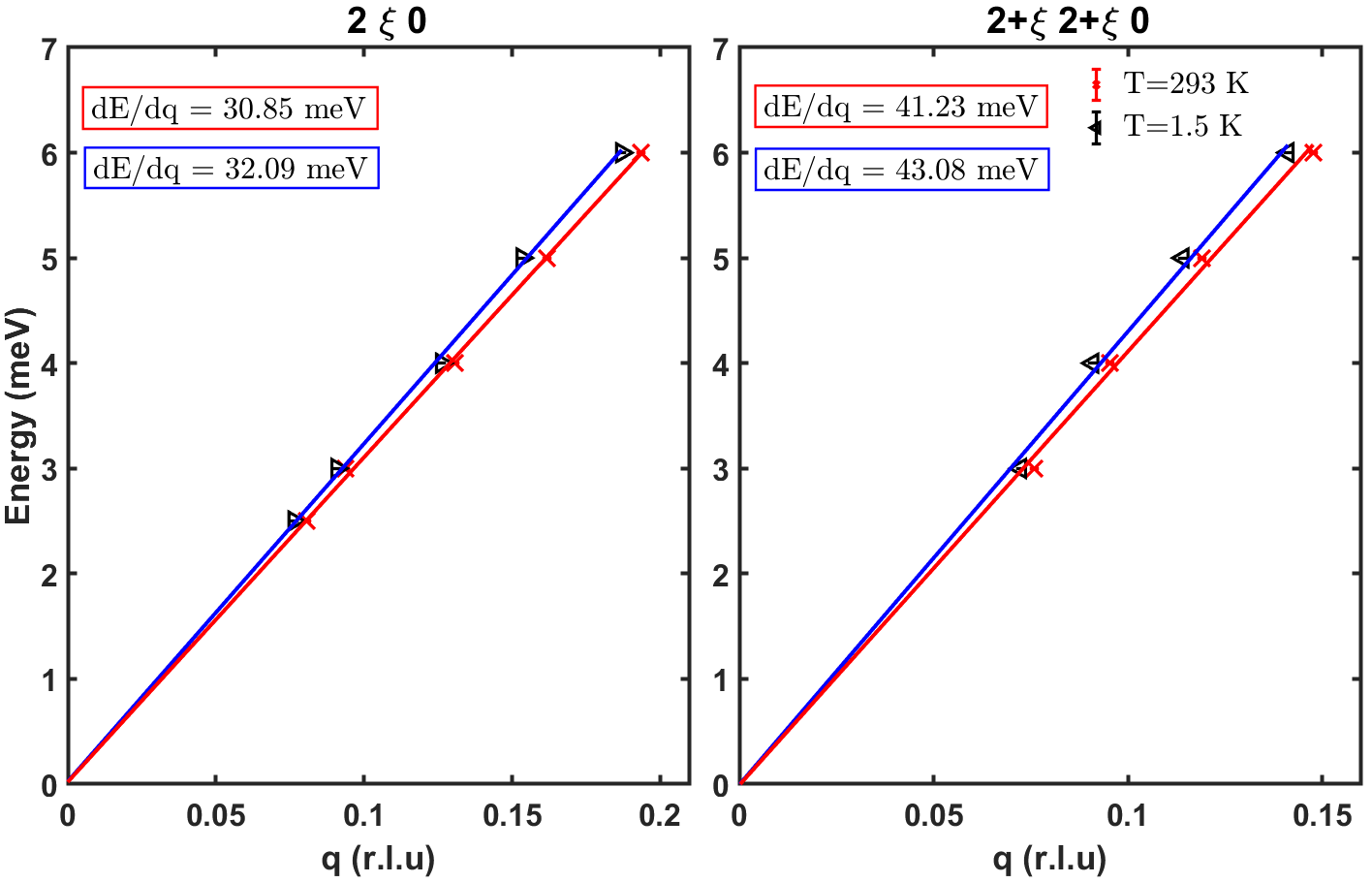}
		\caption{Dispersion obtained from the constant-$E$ scans in Fig.\,\ref{const_energy} using the focusing direction. The scans have been corrected for resolution effects taking the temperature into account $GENAX$. The blue and red lines denote the linear fit to the datasets at 1.6 and 290\,K, respectively. All energy shifts indicate hardening of phonons to about 4\%}
		\label{const_energy_dispersion}
	\end{center}
\end{figure}

\subsection{Absence of nematic instability}

In many FeAs-based materials the nematic transition occurs and is associated with the ferroelastic symmetry reduction from tetragonal to orthorhombic~\cite{Boehmer2022}. In consequence the phase transition as well as nematic precursors in the high-temperature phase couple to the shear modulus $C_{66}$ that softens at the transition.
This elastic constant determines the slope of the transversal acoustic phonons of $\Delta_3$ and $\Sigma_3$ symmetry in our notation. Following the slope of these branches, one can detect an impact of nematic fluctuations as a softening near
the Brillouin zone center and even deduce the nematic correlation length~\cite{Weber2018,Merritt2020}. For LiFeAs only a single mode at 2\,meV was studied but did not show any evidence for softening upon cooling which indicated
the absence of nematic fluctuations in this material~\cite{Merritt2016}.

Figure~\ref{const_energy} shows constant-energy scans along (2 $\xi$ 0) and (2+$\xi$ 2-$\xi$ 0). 
Extra intensity is observed at the center for the low-energy scans, which stems from the contribution of longitudinal modes due to the finite energy and momentum resolution of the IN8 instrument. To correct for this effect and for the shift of maxima due to the asymmetric weighting of the phonon scattering, one needs to convolute the scattering function with the instrument resolution function in order to simulate scans accordingly. 
As it is explained in more detail in the APPENDIX, the resolution corrections can be obtained with the lattice-dynamics model taking all relevant phonon modes into account. For each scan the correction factor was determined with the $GENAX$ program and the data were corrected. The dispersion in Fig.\,\ref{const_energy_dispersion} presents the resolution-corrected data that follow a linear dispersion relation.
The linear fits to the data immediately reveal the normal hardening of the phonon group-velocity while nematic correlations should imply softening. In accordance with the previous analyses \cite{Wissman2022,Merritt2016} we may thus exclude the presence of significant nematic fluctuations in LiFeAs (assuming a similar coupling between electronic nematicity and the lattice as in the other FeAs families).

\section{Conclusion}

The combination of INS experiments with force-constant lattice-dynamics model and DFT calculations yields the full phonon dispersion of the unconventional superconductor LiFeAs with phonon modes separated according to the irreducible representations along main symmetry directions. LiFeAs does not exhibit the structural and magnetic transitions observed in many pure parent compounds of Fe-based superconductors but the layered character of the crystal structure is reflected in a strongly anisotropic thermal expansion. Overall we find good agreement between DFT calculations and the experimental 
phonon dispersion, which indicates that the essential couplings are properly captured in the DFT. In particular we do not find evidence for strong phonon renormalization in the experimental data that could be expected for some
particular strong electron-phonon coupling that is overlooked in the electronic structure model. 

Although some branches considerably soften in the Brillouin zone causing anticrossing behavior, the zone-boundary energies are above $\sim$5\,meV for the $Z$ point (0 0 0.5) and above 12\,meV for $X$ and $M$. The low $Z$ mode energy is just another consequence of the
layered structural character. Note however, that the slopes of longitudinal acoustic branches are similar for in- and 
out-of-plane propagation vectors. The dispersion of branches starting at the infrared-active $u$ modes does not exhibit a steep dispersion near the zone center which would indicate an ionic character with poorly screened Coulomb potentials. 
The in-plane polarized low-energy transversal acoustic phonons with in-plane propagation vectors  exhibit normal behavior and thus confirm the absence of strong nematic fluctuations. We find in general a hardening of phonons upon cooling and a reduction of the line-width that is expected for the usual anharmonic relation between phonon frequencies and the unit-cell volume. But the effects strongly vary between  different modes. This is attributed to the peculiar thermal expansion and the strongest observed hardening of the lowest $E_g$ mode amounts to 6.5\,\%. At the superconducting transition we do not observe any significant change in the phonon energies or line widths.

Overall the phonon dispersion in LiFeAs appears to be rather normal without indication for strong electron-phonon coupling or for a structural instability.

\begin{figure}[htp]
	\centering
	\includegraphics[width=0.66\columnwidth]{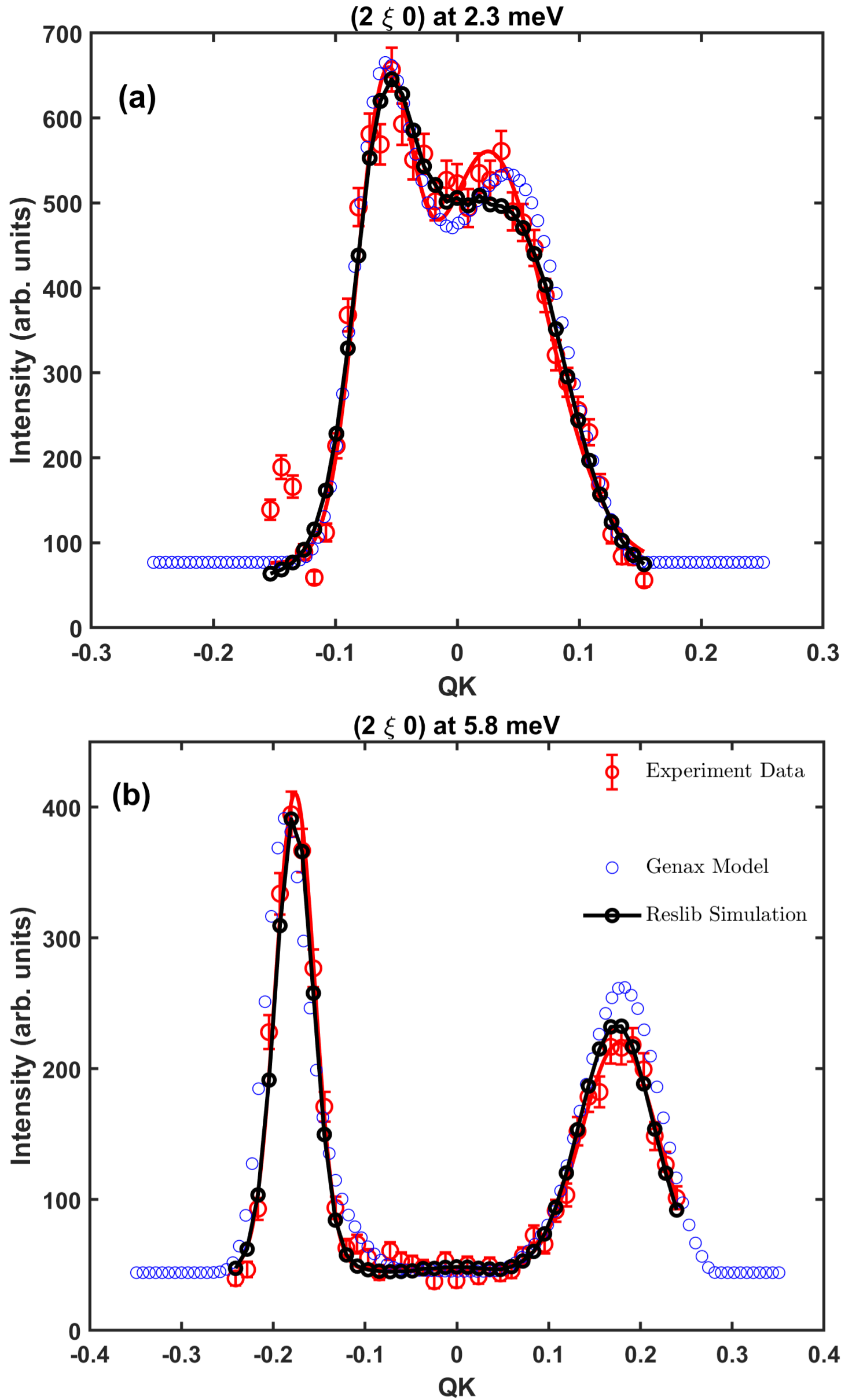}
	\caption{Comparison of measured constant-energy scans across the low-energy transversal in-plane polarized acoustic phonons compared to the simulation with the $RESLIB$ and $GENAX$ packages. The 2.5\,meV results in panel (a) clearly indicate an additional contribution in the scan center that stems from longitudinal modes that contribute due to the
		large resolution ellipsoid in $\bf Q$ space. Here only the $GENAX$ package that folds the entire phonon dispersion
		can properly reproduce the data. The scan at higher energy shown in panel (b) suffers less from this calculation and can be correctly be simulated by both methods.}
	\label{simulation}
\end{figure}

\section{Appendix}

- {\it Resolution correction for acooustic modes} - The observed INS intensity along a scan path results from the folding of the scattering function $S(\bf{Q},E)$ with the 
fourdimensional (4D) resolution function of the TAS. In most cases one may correctly determine the mode energy
in a constant-$\bf Q$ scan or the propagation vector in a constant-$E$ scan by fitting a Gaussian distribution.
However, at low energies (where $q$$\rightarrow$0) the Bose and the $1/\omega$ terms in the dynamical structure factor result in a strong asymmetry of the scattering function (see Eq.\,(\ref{dynamical-struc-fac})) so that the folded intensity distribution does not peak at the true phonon energy or phonon propagation vector. This resolution-induced shift can be estimated by simulating a phonon scan and fitting the calculated intensity distribution with a Gaussian distribution. The difference between the phonon position of the model and the fitted maximum is the resolution induced shift that then can be used to correct the fit values with the experimental data. Of course the phonon dispersion and the simulated scattering strength of the model should be as close as possible to the true dispersion of the system.

\begin{figure*}[th]
	\begin{center}
		\includegraphics[width=0.85\textwidth]{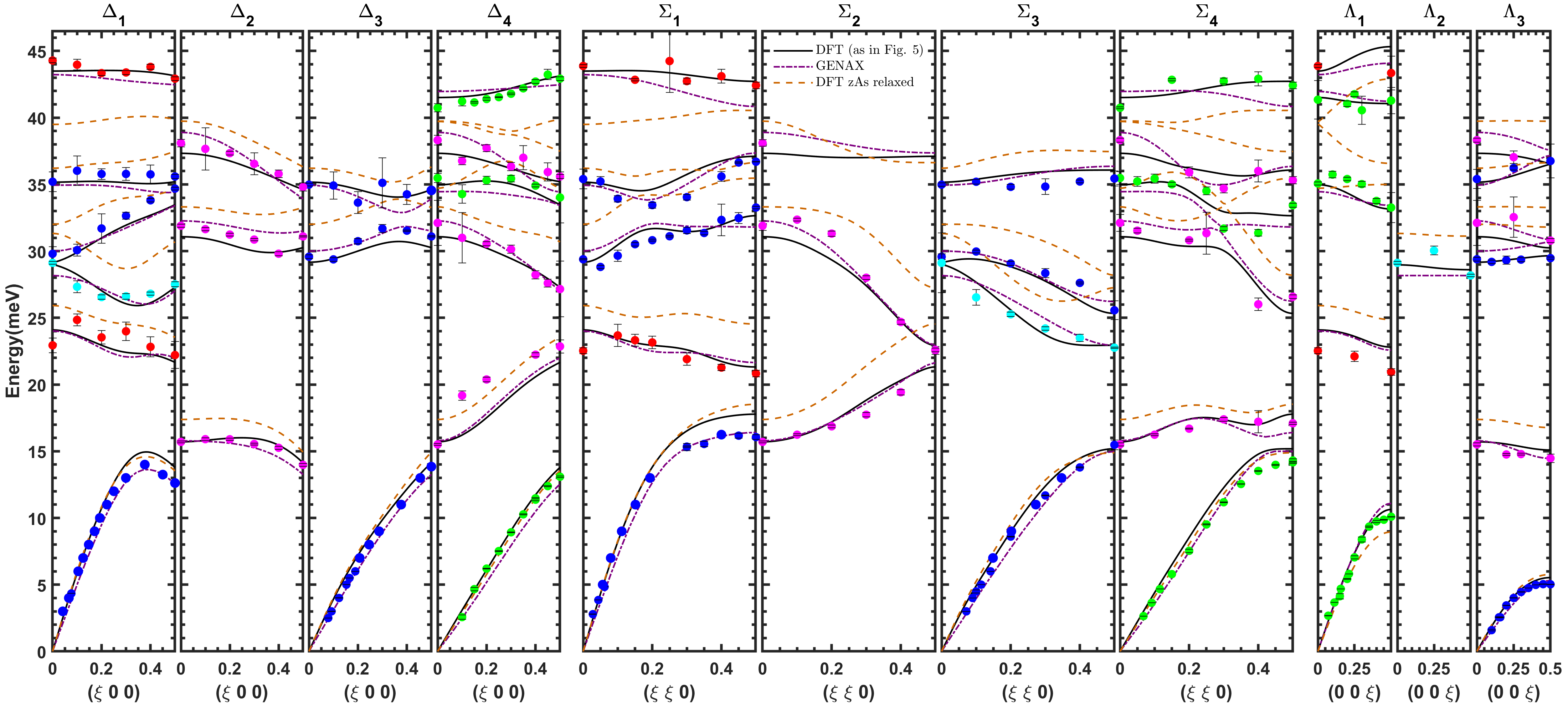}
		\caption{ Phonon dispersion in LiFeAs. Symbols present the experimental energy values with the same color coding as in Fig.~\ref{phonondispersion}. Black solid lines and orange dashed lines denote the DFT calculation discussed in section V and VI and those with the optimized As-$z$ position; the magenta broken lines show the results of the force-constant model.}
		\label{GENAX_VS_data}
	\end{center}
\end{figure*}

We used the $GENAX$~\cite{genax} and $RESLIB$~\citep{Zheludev2006} program packages to simulate the folding.
$RESLIB$ analytically calculates the 4D resolution ellipsoid in momentum-energy space using Cooper-Nathans or Popovici approximations~\cite{Shirane,Zheludev2006}, and $GENAX$ uses the Cooper-Nathans formalism with effective mosaic values and beam divergences to approximate the focusing.
The resolution matrix $\bs M$ (a 4x4 matrix) describes the Gaussian broadening:

\begin{equation}
	R(\Delta Q_x, \Delta Q_y, \Delta Q_z, \Delta E) \propto \exp\left( -\frac{1}{2} \vec{v}^T \bs M \vec{v} \right),
	\end{equation}

where $\vec{v}$\,=\,$(\Delta Q_x, \Delta Q_y, \Delta Q_z, \Delta E)$ is the distance vector in the 4D space. This resolution function is convoluted with the theoretical scattering function to simulate the observed scans.

With the $RESLIB$ package the scattering function arising from transversal phonons is simulated by a linear
dispersion $E(q)$ combined with a one-phonon scattering strength that assumes a perfect acoustic polarization and thus only incorporates the trivial Bose and the $1/\omega$ terms by adapting the slope of the phonon branch. With the full lattice-dynamics calculations provided by the $GENAX$ program package one can calculate the complete scattering function at any $\bf q$ value. At high energy both methods
yield comparable results, but at low energy the contribution of longitudinal modes that are captured by the finite $\bf Q$ resolution yield additional scattering near the $\xi$=0 centers of the constant-E scans that only the full analysis can properly describe, see Fig.\,\ref{simulation}.

- {\it Comparison with force-constant model} - In Fig.~\ref{GENAX_VS_data} we compare the experimental phonon dispersion with the results of the refined  force-constant model. With the limited amount of 24 force constants the essential features of the dispersion, in particular the anticrossing behaviors are well described. Therefore, also the identification of the individual modes through the dynamical structure factor obtained from the polarization vectors appears reliable.

- {\it Dependence of DFT calculations on structure parameters} - It is well known that the As-$z$ parameter, which determines the distance of As from the Fe plane, is an important structural parameter for the 1111 and 122 Fe-pnictide families, and has a large influence on the positions of electronic bands near the Fermi energy. It has been noticed in early DFT studies, that the theoretically optimized As-$z$ value deviates significantly from the experimental one \cite{Yin2008,Mazin2008b}. Furthermore, an intriguing relationship between the As position and the ordered magnetic moment of Fe was found, and the agreement with experiment improved when the magnetic order was taken into account \cite{Yin2008,Yildrim2009}. This peculiar magneto-structural coupling also manifests itself in lattice-dynamical properties. In theoretical studies of BaFe$_2$As$_2$, certain phonons react sensitively on the As position and magnetic order, and better agreement with measured phonon energies is observed for magnetic calculations \cite{Reznik2009}. A similar trend was found for LaFeAsO \cite{Hahn2013}.

An alternative explanation was suggested by previous DFT+DMFT studies, which provide an enhanced correlation treatment. Here, an improved As position was found even without considering magnetic order \cite{Aichhorn2011}. This points to the possibility that the wrong prediction of the As position within standard DFT might be due to an insufficient treatment of correlation.

For LiFeAs we do observe very similar trends. As discussed in Sec. V and VI, the good agreement of the calculated phonon dispersion with measurement (see Fig.\,\ref{phonondispersion}) is obtained for a lattice geometry, where the As-$z$ parameter is taken from experiment. Figure~\ref{GENAX_VS_data} compares this dispersion with a calculation, where the theoretically optimized value is used instead. However, this calculation results in poor agreement. Significant discrepancies for optical branches in the middle part of the spectrum as well as softening of the highest branches highlight the sensitivity of the phonon spectrum on the As position. Because LiFeAs does not exhibit magnetic order, the discrepancy cannot simply be attributed to the neglect of magnetic order. Future work has to show, if the wrong DFT prediction for the As position can be reconciled by including local magnetic moments (paramagnetic state) or by an improved treatment of correlations.

%
%\begin{figure}[htp]
%	\centering
%	\includegraphics[width=0.94\columnwidth]{opticalmodes-FWHM.png}
%	\caption{Resolution corrections implemented via $RESLIB$ and $GENAX$}
%	\label{simulation}
%\end{figure}

\begin{acknowledgments}
	AT and MB acknowledge support by the Deutsche Forschungsgemeinschaft (DFG, German Research Foundation) through  project number 461247437 BR 2211/3-1. SW acknowledges support by the DFG through  project number 461247437 WU595/17-1. RH acknowledges support by the state of Baden-W\"urttemberg through bwHPC.
\end{acknowledgments}

\begin{center}
	{\bf DATA AVAILABILITY}
\end{center} 

The data that support the findings of this article are openly 
available \cite{zenodo}.

%\bibliography{Fe-SC_n}
%Control: production of eprint (1) enabled
%

\end{document}